\documentclass[12pt]{iopart}

\usepackage{cite}

\usepackage{graphicx}
\usepackage{dcolumn}
\usepackage{bm}
\expandafter\let\csname equation*\endcsname\relax
\expandafter\let\csname endequation*\endcsname\relax
\usepackage{amsmath}
\usepackage{amssymb}
\usepackage{physics}
\usepackage{braket}
\usepackage[dvipsnames]{xcolor}
\usepackage[style=base]{caption}
\usepackage[group-separator={,},group-minimum-digits=4]{siunitx}
\DeclareSIUnit{\au}{a.u.}
\ExplSyntaxOn
\NewDocumentCommand{\au}{m}
 {
  \SI{#1}{{a.u.}}
  \peek_charcode_remove:NT .
   {
    \mode_if_math:F { \spacefactor\sfcode`\.\scan_stop: }
   }
 }
\ExplSyntaxOff
\usepackage{hyperref}
\usepackage{cleveref}

\hypersetup{hypertexnames=false}
\usepackage[section]{placeins}
\usepackage[numbers]{natbib}
\usepackage[outdir=./]{epstopdf}

\usepackage{floatrow}
\usepackage[labelformat=simple,label font={normalsize}]{subfig}            

\renewcommand{\v}[1]{\bm{#1}} 
\renewcommand{\u}[1]{\bm{\hat{#1}}}  
\newcommand{\Fe}[1]{Fe\textsubscript{#1}}
\newcommand*\mean[1]{\overline{#1}} 

\hypersetup{
	colorlinks = true,
	urlcolor={blue},
	linkcolor={blue},
	citecolor={blue}
}

\begin{document}

\title[The Einstein-de Haas Effect in an Fe\textsubscript{15} Cluster]{The Einstein-de Haas Effect in an Fe\textsubscript{15} Cluster}


\author{T Wells$^1$, W M C Foulkes$^2$,
S L Dudarev$^3$ and A P Horsfield$^1$}
\address{$^1$ Department of Materials and Thomas Young Centre, Imperial College London, South Kensington Campus, London SW7 2AZ, United Kingdom}
\address{$^2$ Department of Physics and Thomas Young Centre, Imperial College London, South Kensington Campus, London SW7 2AZ, United Kingdom}%
\address{$^3$ UK Atomic Energy Authority, Culham Center for Fusion Energy, Oxfordshire OX14 3DB, United Kingdom}
\address{$^3$ Department of Physics and Thomas Young Centre, Imperial College London, South Kensington Campus, London SW7 2AZ, United Kingdom}%
\ead{\mailto{tomos.wells11@imperial.ac.uk}}

\vspace{10pt}
\begin{indented}
\item[]\today
\end{indented}

\begin{abstract}
Classical models of spin-lattice coupling are at present unable to accurately reproduce results for numerous properties of ferromagnetic materials, such as heat transport coefficients or the sudden collapse of the magnetic moment in hcp-Fe under pressure. This inability has been attributed to the absence of a proper treatment of effects that are inherently quantum mechanical in nature, notably spin-orbit coupling. This paper introduces a time-dependent, non-collinear tight binding model, complete with spin-orbit coupling and vector Stoner exchange terms, that is capable of simulating the Einstein-de Haas effect in a ferromagnetic \Fe{15} cluster. 
The tight binding model is used to investigate the adiabaticity timescales that determine the response of the orbital and spin angular momenta to a rotating, externally applied $B$ field, and we show that the qualitative behaviours of our simulations can be extrapolated to realistic timescales by use of the adiabatic theorem. 
An analysis of the trends in the torque contributions with respect to the field strength demonstrates that SOC is necessary to observe a transfer of angular momentum from the electrons to the nuclei at experimentally realistic $B$ fields.
The simulations presented in this paper demonstrate the Einstein-de Haas effect from first principles using a Fe cluster.
\end{abstract}

%
%
%
%
%

\section{\label{sec:introduction}Introduction}


Developing materials for use close to the plasma in a tokamak, where the heat and neutron fluxes are high, is a challenge as few solids survive undamaged for long~\cite{STORK2014277}. Iron-based steels are proposed as structural materials for blanket modules and structural components due to their ability to withstand intense neutron irradiation. At the same time, there are no reliable data about the performance of steels under irradiation in the presence of magnetic fields approaching 10~T. Despite the durability of steels, however, the service lifetimes of reactor components are limited. This has led to a resurgence of research into the properties of steels and other materials under reactor conditions. Properties of interest include thermal conductivity coefficients, and other physical and mechanical properties, intimately related to the question about how the heat and radiation fluxes affect the microstructure of reactor materials~\cite{Alshits}. Another focus is on optimizing the structural design to improve the tritium breeding ratio, the heat flow, and structural stability of reactor components~\cite{KONCAR2017567,Dudarev2018,Reali2022}.

An example of an improvement based on research into the properties of ferromagnetic materials is as follows. Austentitic Fe-Cr-Ni steels are used in the ITER tokamak~\cite{0029-5515-41-3-302}, and are non-magnetic on the macroscopic scale while being microscopically antiferromagnetic. In the next generation demonstration fusion reactor (DEMO), however, the blanket modules are expected to be manufactured from ferromagnetic ferritic-martensitic steels, because these have been found to exhibit superior resistance to radiation damage~\cite{STORK2014277,KOHYAMA}.

Given the importance of ferritic steels in reactor design, it would be helpful to understand heat flow in iron in the presence of strong magnetic fields at high temperature. This is a difficult task. A good heat-flow model must reproduce the dynamics of a many-atom system with complicated inter-atomic forces, whilst also describing the electronic thermal conductivity and the influence of spin-lattice interactions. The spins are not all aligned above the Curie temperature, but they are still present and still scatter electrons. The exchange interactions between spins also affect the forces on the nuclei. Our aim in this paper is to begin the development of such a model, starting from the quantum mechanical principles required to understand electrons and spins.




Although a full treatment of the behaviour of the spins and electrons requires quantum theory, various classical atomistic models have been established to investigate spin-lattice interactions. In 1996, Beaurepaire et al.\ developed the three temperature model (3TM)~\cite{PhysRevLett.76.4250}, a nonequilibrium thermodynamics-based approach to describe the interactions between the lattice, spin, and electronic subsystems. A microscopic 3TM proposed by Koopmans et al.\ in 2010 was able to explain the demagnetization timescales in pulsed-laser-induced quenching of ferromagnetic ordering across three orders of magnitude~\cite{Koopmans2010}.

Langevin spin dynamics (SD), developed in~\cite{PhysRev.130.1677,PhysRevB.58.14937,doi:10.1063/1.365023,Lyberatos_1993,doi:10.1063/1.352594,TSIANTOS2002999,CHUBYKALO200328,1233125,PhysRevB.83.134418}, builds on classical molecular dynamics by adding fluctuation and dissipation terms to the equations of motion for the particles and their spins. Its most well known application is simulating relaxation and equilibration processes in magnetic materials at finite temperatures~\cite{PhysRevB.58.14937,CHUBYKALO200328,Radu2011}.

In 2008, Ma et al.\ \cite{Ma2008} used a model in which atoms interact via scalar many-body forces as well as via spin orientation dependent forces of the Heisenberg form to predict isothermal magnetization curves, obtaining good agreement with experiment over a broad range of temperatures. Further, they showed that short-ranged spin fluctuations contribute to the thermal expansion of the material.

In 2012, Ma et al.\ \cite{Ma2012} proposed a generalized Langevin spin dynamics (GLSD) algorithm that builds on Langevin SD by treating both the transverse (rotational) and longitudinal degrees of freedom of the atomic magnetic moments as dynamical variables. This allows the magnitudes of the magnetic moments to vary along with their directions.  The GLSD approach was used to evaluate the equilibrium value of the energy, the specific heat, and the distribution of the magnitudes of the magnetic moments, and to explore the dynamics of spin thermalization.

In 2022, Dednam et al.\ \cite{DEDNAM2022111359} used the spin-lattice dynamics code, SPILADY, to carry out simulations of Einstein–de Haas effect for a Fe nanocluster with more than 500 atoms. Using the code, the authors were able to show that the rate of angular momentum transfer between spin and lattice is proportional to the strength of the magnetic anisotropy interaction, and that full spin-lattice relaxation was achievable on 100 ps timescales.

Despite the efforts invested in classical models of spin-lattice interactions, they exhibit numerous shortcomings, such as the inability to accurately reproduce the measured heat transport coefficients in ferromagnetic materials~\cite{Duffy_2006,Race_2009,Ma2012}, or the sudden collapse of the magnetic moment in hcp-Fe under pressure, which is thought to be a consequence of spin-orbit coupling (SOC)~\cite{Iota2007}. These limitations can only be overcome by switching to a quantum mechanical description.

Experimental work on isolated clusters has improved the understanding of the differences in magnetic properties between atomic and bulk values.
Stern-Gerlach experiments have been used to study the magnetic moment per atom of isolated clusters, as a function of the external magnetic field and temperature. These experiments found that the average magnetization as a function of field strength and temperature, resembles the Langevin function, which was initially attributed to thermodynamic relaxation of the spin while in the magnetic field~\cite{Khanna1991,Bucher1991,Douglass1993}. Using a model based on avoided crossings between coupled rotational and spin degrees of freedom, Xu et al. subsequently explained why the average magnetization resembles the Langevin function for all cluster sizes, including for low temperatures, without reference to the spin-relaxation model~\cite{Xu2005}. 





Addressing the physics of iron out of equilibrium --- a complicated time-evolving system of interacting nuclei, electrons and spins --- in a quantum mechanical framework is such a challenge that we seek first to understand one of the simplest phenomena involving spin-lattice coupling: the Einstein-de Haas (EdH) effect. The EdH effect is, of course, the canonical example of how electronic spins apply forces and torques to a crystal lattice, and has been well studied experimentally. It is perhaps surprising, therefore, that we were unable to find any published quantum mechanical simulations of the EdH effect for bulk materials. This paper builds on previous work, in which we reported simulations of the EdH effect for a single O\textsubscript{2} dimer~\cite{twells}.


In 1908, O.\ W.\ Richardson was the first to consider the transfer of angular momentum from the internal ``rotation'' of electrons (i.e., the magnetic moments within the material) to the mechanical rotation of macroscopic objects~\cite{Richardson1908}. Inspired by Richardson’s paper, S.\ J.\ Barnett theorized the converse effect, in which the mechanical rotation of a solid changes the magnetic moments~\cite{Barnett1909}. Many experiments sought to measure the ratio $\lambda = \frac{\Delta \v J}{\Delta \v M}$, where $\Delta \v J$ is the change in electronic angular momentum and $\Delta \v M$ is the change in magnetization of the material. Due to the lack of understanding of electron spin at the time, Richardson and Barnett predicted that $\lambda$ should equal $\frac{e}{2m}$; the true value is closer to $\frac{e}{m}$.

The EdH effect was named after the authors of the 1915 paper \cite{EdH} that reported the first experimental observations, finding $\lambda = \frac{e}{2m}$ to within the measurement uncertainty. In the same year, Barnett published the first observations of the Barnett effect \cite{Barnett1915}, with a more accurate measurement closer to $\lambda = \frac{e}{m}$. Subsequent measurements of the Einstein-de Haas effect by Stewart~\cite{Stewart1918} supported the result $\lambda = \frac{e}{m}$. The discrepancy between the predicted and measured values of $\lambda$ came to be known as the gyromagnetic anomaly, and was finally resolved only after it was understood that most of the magnetization can be attributed to the polarization of the electrons' spins~\cite{Reck1969}.

The ferromagnetic resonance of Larmor precession observed in 1946 by Griffiths~\cite{GRIFFITHS1946} provided a more accurate technique for measuring gyromagnetic ratios, superseding measurement of the EdH and Barnett effects. Using ferromagnetic resonance, Scott accurately measured the gyroscopic ratios of a range of ferromagnetic elements and alloys~\cite{RevModPhys.34.102}. After this point, interest in the EdH effect reduced as it was widely considered to be understood.


In this paper, we describe the implementation of a non-collinear tight-binding (TB) model complete with all features required to capture spin-lattice coupling in iron in the presence of a time-dependent applied magnetic field. The required features are: coupling of the electrons to the lattice, coupling of the electron magnetic dipole moment to an externally applied time-dependent magnetic field, coupling between orbital and spin angular momentum through SOC, and electron exchange. Using this model, we simulate the response of an Fe\textsubscript{15} cluster to a time-varying magnetic field, and analyse the torque on the nuclei due to the electrons.
We find that, in a slowly rotating $B$ field, the orbital and spin angular momenta rotate with the field, leading to a measurable torque on the cluster. We also describe the qualitative features of the evolution of the spin and angular momentum, and demonstrate the enhancement of the torque exerted on the nuclei by the electrons as a result of SOC. Thus, this work documents a quantum mechanical model capable of simulating the Einstein-de Haas effect, and reveals the physical mechanisms that set the timescales over which the spins evolve. 

This paper is structured as follows. Section~\ref{sec:theory} describes the method used for the calculations. The results of the simulations are discussed in Sec.~\ref{sec:results}. Conclusions are drawn in Sec.~\ref{sec:conclusions}.

\section{\label{sec:theory}Theory}

\subsection{\label{sec:system}The System}

\begin{figure}[htp!]
	\centering
	\includegraphics[width=0.5\linewidth]{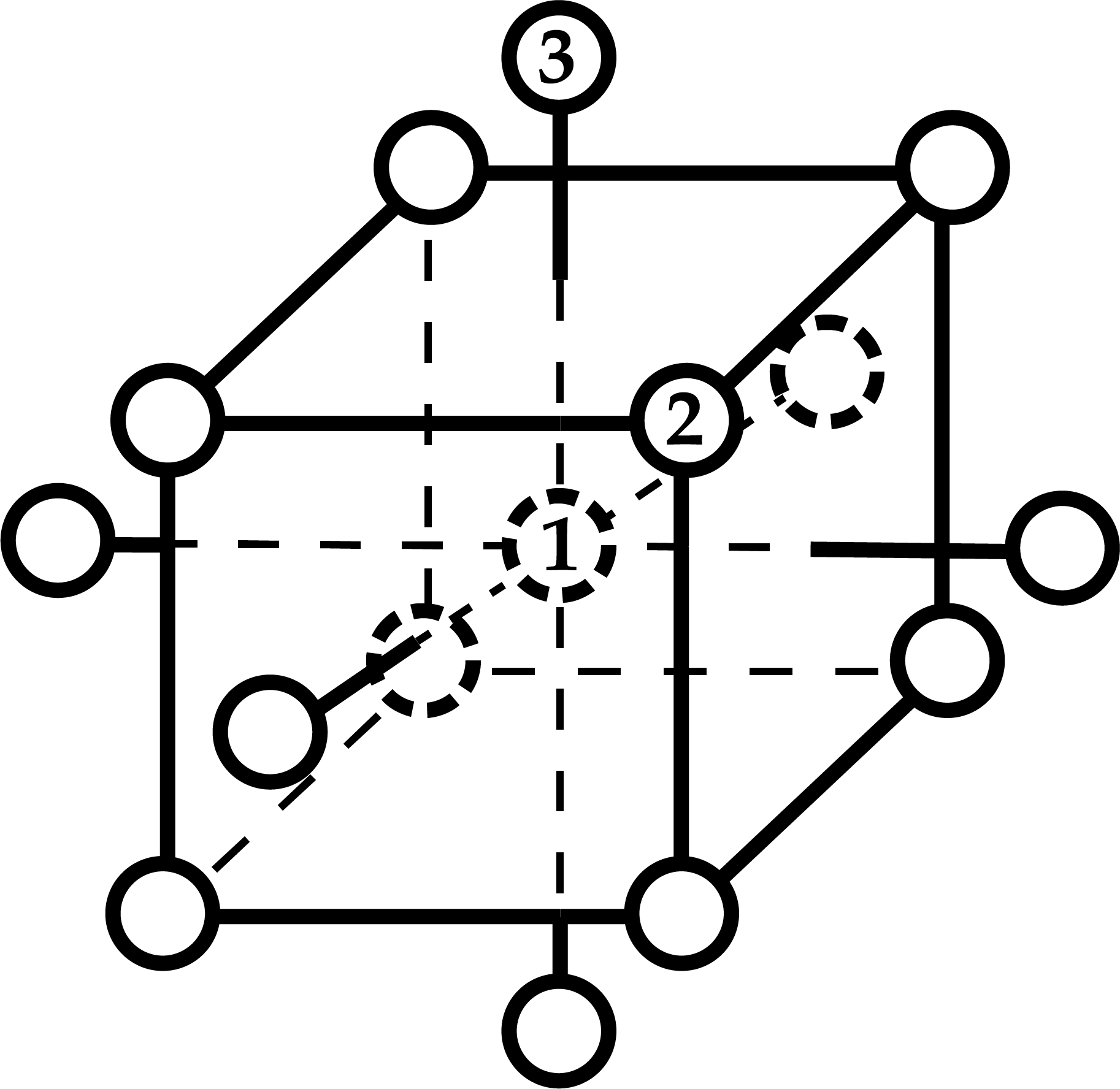}
	\caption{The \Fe{15} cluster chosen for analysis in this work. The eight type-2 atoms are the nearest neighbours of atom 1 and the six type-3 atoms are the next-to-nearest neighbours. The nearest-neighbour distance is $\SI{2.49}{\angstrom}$.}
	\label{fig:Fe15}
\end{figure}

The Fe cluster studied in this work is Fe\textsubscript{15}, in the configuration shown in figure~\ref{fig:Fe15}. This cluster has been studied numerically in many previous works~\cite{yang1981,vega1993,franco1999,Sipr2004,JO20011045,Kohler2005}. The 15 atoms are positioned exactly as in a subset of the body-centered cubic (BCC) lattice, with the nearest-neighbor distance set to $\SI{2.49}{\angstrom}$ to match that of bulk iron~\cite{yang1981}. The atoms in the cluster are held in position and are not allowed to move during the simulation.

The TB basis functions are atomic-like $d$ orbitals (using real cubic harmonics) with separate orbitals for up and down spins to form a non-collinear TB model. The ten basis functions on each atom are denoted,
\begin{align}
	\ket{d_{z^2,\uparrow}}, ~ \ket{d_{xz,\uparrow}}, ~ \ket{d_{yz,\uparrow}}, ~ \ket{d_{xy,\uparrow}}, ~ \ket{d_{x^2-y^2,\uparrow}}, \notag\\
	\ket{d_{z^2,\downarrow}}, ~ \ket{d_{xz,\downarrow}}, ~ \ket{d_{yz,\downarrow}},  ~ \ket{d_{xy,\downarrow}}, ~  \ket{d_{x^2-y^2,\downarrow}}.
\end{align}
The basis set does not include any $s$ or $p$ orbitals below the $3d$ shell or any orbitals above it. We use the TB model of Liu \textit{et~al.}~\cite{OxfordModel}, hereafter called the Oxford TB model. This model for Fe has 6.8 electrons in the $3d$ shell. With 15 Fe atoms, the full Hamiltonian is a $150\times150$ Hermitian matrix and the 150 molecular orbitals (MOs) are occupied by 102 electrons. The non-magnetic terms in our Hamiltonian matrix are exactly as described in~\cite{OxfordModel}, and do not introduce any terms that couple spin to the lattice degrees of freedom; the magnetic, spin-orbit and exchange terms are discussed in Sec.~\ref{sec:Hamiltonian}.

The MOs $\ket{\phi_n}$ are the eigenfunctions of the Hamiltonian and can be expressed as linear combinations of the basis states $\ket{\chi_{\alpha\sigma}}$, assumed to be orthonormal, with expansion coefficients $d_{n\alpha\sigma}$,
\begin{align}
	\ket{\phi_n} = \sum_{\alpha\sigma} d_{n\alpha\sigma}\ket{\chi_{\alpha\sigma}},
\end{align}
where $\alpha$ runs over the spatial atomic orbitals (AOs) on all atoms and $\sigma$ is a spin index taking the values up ($\uparrow$) or down ($\downarrow$).

\subsection{The simulations}

The $B$ field produced by a fixed solenoid reverses its direction when the current reverses, remaining parallel or anti-parallel to the solenoidal axis but changing in magnitude. In our simulations, however, we chose to study a rotating $B$ field of constant magnitude: the $B$ vector traces out a semicircle, from the south pole ($-\u z$) to the north pole ($+\u z$) of a sphere.

There are three reasons we believe that this rotational path better mimics the field experienced by a single magnetic domain in a measurement of the EdH effect. (i) The magnetic field felt by a single magnetic domain within a solid is not in general exactly aligned with its magnetization axis due to the configuration of the other surrounding domains. This symmetry-breaking mechanism is absent when a field with fixed direction is applied to a single domain, the magnetization of which is initially aligned with the applied field. (ii) It is unlikely that the crystal lattice of any single magnetic domain is aligned such that the initial and final fields are exactly parallel to the easy axes of the domain. (iii) The total exchange energy is large even for a small cluster and scales with system size. For the magnetization of an isolated single-domain cluster to reverse its direction in response to a $B$ field that is initially aligned with the magnetization and reverses along its axis, the Stoner moment would have to pass through zero, overcoming a large exchange energy barrier. We deem this scenario unlikely. In a real multi-domain magnet, the spin stays large and rotates rather than passing through $\expval{\v S} = \v 0$. 

To perform the rotation from the south pole to the north pole of a sphere, the $\v B$ field is parametrized in spherical coordinates as
\begin{align}
	\v B = (-B\sin\theta, \; 0, \; -B\cos\theta),
\end{align}
where $\theta = \omega t$, $t$ is the elapsed time, $\omega = \pi / T_f$, and $T_f$ is the time at which the simulation finishes. The magnitude $B = \lvert \v B \rvert$ of the applied magnetic field differs in different simulations. The $\v B$ field is initially in the $-{\u z}$ direction and gradually rotates by $180^\circ$ in the $xz$ plane. The simulation is complete when $\v B$ points in the $+{\u z}$ direction.


\subsection{The Time Evolution Algorithm}

At the beginning of the simulation ($t=0$), the molecular orbitals $\ket{\phi_n}$ are obtained by diagonalizing the self-consistent ground state Hamiltonian. At later times, the state is calculated from the time-evolved molecular orbitals $\ket{\psi_n(t)}$, which are found by solving the time-dependent Schr\"{o}dinger equation,
\begin{align}
    i\hbar\partial_t \ket{\psi_n(t)} = H(t) \ket{\psi_n(t)},
\end{align}
subject to the initial condition $\ket{\psi_n(t=0)} = \ket{\phi_n}$. For times later than $t=0$, the time-evolved molecular orbitals are not exact eigenfunctions of the Hamiltonian, since the Hamiltonian $H(t)$ depends on time if $\v B(t)$ depends on time.

The time-dependent expansion coefficients $d_{n\alpha\sigma}(t)$ are defined by
\begin{align}
    \ket{\psi_n(t)} = \sum_{\alpha\sigma} d_{n\alpha\sigma}(t) \ket{\chi_{\alpha\sigma}}, \label{eq:expansion_coefficients}
\end{align}
and satisfy the discrete equivalent of the time-dependent Schr\"{o}dinger equation,
\begin{align}
    i\hbar\frac{\partial}{\partial t} d_{n\alpha\sigma}(t) = \sum_{\alpha'\sigma'} H_{{\alpha\sigma},{\alpha'\sigma'}}(t) d_{n{\alpha'\sigma'}}(t).
\end{align}
Rewriting this equation of motion in matrix-vector form with $(\v d)_{n\alpha\sigma} = d_{n\alpha\sigma}$ and $(\v H)_{{\alpha\sigma},{\alpha'\sigma'}} = H_{{\alpha\sigma},{\alpha'\sigma'}}$, gives
\begin{align}
    i\hbar \frac{\partial \v d(t)}{\partial t} = \v H(t) \v d(t).
    \label{eq:Devolution}
\end{align}
To solve Eq.~(\ref{eq:Devolution}) numerically, we introduce a small but finite positive time step $\delta t$ and use the finite-difference approximation~\cite{Hatano2005}
\begin{align}
    \v d(t+\delta t) = \textrm{exp}\bigg(\frac{\v H(t+\frac{1}{2}\delta t)}{i\hbar} \delta t \bigg) \v d(t),
\end{align}
which is both time-reversible and unitary. In index notation, one step of the time evolution takes the form
\begin{align}
    d_{n\alpha\sigma}(t+\delta t) = \sum_{\alpha'\sigma'}\Big(e^{H(t+\frac{1}{2}\delta t) \delta t / i\hbar}\Big)_{{\alpha\sigma},{\alpha'\sigma'}} d_{n\alpha'\sigma'}(t). \label{eq:timestep}
\end{align}
The calculation of $H(t+\frac{1}{2}\delta t) \delta t / i\hbar$ from Eq.~\eqref{eq:timestep} is not trivial since the Stoner term must be extrapolated to $t+\frac{1}{2}\delta t$ based on its previous values. The method employed for this task is described in~\cite{twells}.

The initial condition, $\ket{\psi_n(0)} = \ket{\phi_n}$, implies that the coefficients $d_{n\alpha\sigma}$ at $t=0$ are given by
\begin{align}
    d_{n\alpha\sigma}(0) = \braket{\chi_{\alpha\sigma}|\phi_n}.
\end{align}

The time-dependent one-particle density operator, $\rho(t)$, is defined by
\begin{equation}
    \rho(t) = \sum_{n\;\textrm{occ}} \ket{\psi_n(t) } \bra{ \psi_n(t) }.
\end{equation}
Using Eq.~\eqref{eq:expansion_coefficients}, the matrix elements of $\rho(t)$ may be expressed in terms of the expansion coefficients as
\begin{equation}
	\rho_{\alpha'\sigma',\alpha\sigma}(t) = \sum_{n\;\textrm{occ}} d^{\phantom{*}}_{n\alpha'\sigma'}(t) d^*_{n\alpha\sigma}(t).
	\label{eq:rho_from_d}
\end{equation}

\subsection{The Hamiltonian}\label{sec:Hamiltonian}

To describe the EdH effect, the Hamiltonian must include: (i) coupling of electrons to an external time-dependent magnetic field; (ii) spin-orbit coupling; and (iii) Stoner exchange. In electronic structure methods, Stoner exchange is often used in its collinear form, but this is inappropriate for describing spin dynamics as it breaks rotational symmetry in spin space~\cite{Coury2016}. We therefore use a non-collinear exchange Hamiltonian.

The full Hamiltonian may be partitioned as
\begin{align}
	H = H_0 + H_B + H_{\textrm{SOC}} + H_{\textrm{ex}},\label{eq:hamiltonian}
\end{align}
where $H_0$ is the basic tight-binding Hamiltonian given by the Oxford model, $H_B$ is the interaction with the external field, $H_{\textrm{SOC}}$ describes SOC, and $H_{\textrm{ex}}$ is the vector Stoner exchange term.

The Hamiltonian term that describes the interaction of a single atom with an external magnetic field is
\begin{align}
	H_{B,a}&=-\v{\mu}_a\cdot\v{B}(t) \notag \\
	&= \frac{\mu_B}{\hbar} \sum_{a} P_a (\v{L} + 2 \v{S}) P_a \cdot \v{B}(t) ,
\end{align}
where $\v{\mu}_a$ is the magnetic moment of atom $a$, $P_a=\sum_{m\sigma} |\chi_{am\sigma} \rangle\langle \chi_{am\sigma}|$ is the projection operator on to the basis of atomic-like $d$ orbitals on atom $a$, and $m$ runs over the 5 $d$ orbitals on atom $a$, $\mu_B$ is the Bohr magneton, $\v S$ is the spin angular momentum operator, and $\v L$ is the orbital angular momentum operator about the nucleus in the Coulomb gauge. This form may be justified by reference to the Pauli equation~\cite{cohen-tannoudji}. The magnetic Hamiltonian for the cluster is obtained by summing atomic contributions:
\begin{align}
	H_B &= - \sum_{a} \v{\mu}_{a} \cdot \v{B}(t).
	\label{eq:HB}
\end{align}

The spin-orbit coupling term is of relativistic origin and can be derived by application of the Foldy-Wouthuysen transformation to the Dirac equation~\cite{Foldy1950}. In the spherical potential of a single atom, this gives
\begin{equation}
	H_{\textrm{SOC}} = \frac{1}{2 m_e^2 c^2}\frac{1}{r} \frac{dV(r)}{dr} \v{L}\cdot\v{S},
\end{equation}
where $m_e$ is the mass of an electron, $c$ is the speed of light, and $V(r)$ is the potential experienced by an electron due to the atomic nucleus and the other electrons belonging to that atom within the central field approximation. The radial part of the SOC matrix element between two AOs in the same shell is a constant, $\xi$, and our TB model includes only one shell of AOs per iron atom, so~\cite{LandauQM}
\begin{equation}
	H_{\textrm{SOC}} \approx \frac{\xi}{\hbar^2} \v{L} \cdot \v{S}.
\end{equation}
Since the gradient of the nuclear potential is largest very near to the nucleus, the spin-orbit term can be assumed to couple atomic orbitals on the same atom only. Adding similar terms for every atom in the cluster yields the SOC Hamiltonian used in this work:
\begin{equation}
	H_{\textrm{SOC}} \approx \frac{\xi}{\hbar^2} \sum_{a} (P_a \v{L} P_a) \cdot (P_a \v{S} P_a).
\end{equation}

The Stoner exchange term, which is a mean-field approximation to the many-body effect of exchange, is given by
\begin{align}
	H_\textrm{ex} = - I \sum_a \v m_a \cdot (P_a \v \sigma P_a),
\end{align}
where $I$ is the Stoner parameter (which has units of energy),
\begin{align}
	\v m_a(t) &= \expval{P_a \v \sigma P_a}
\end{align}
is the expectation value of the operator $P_a \v \sigma P_a$,
and $\v \sigma$ is the vector of Pauli matrices. The origin of the Stoner exchange term is described in more detail in~\cite{twells}.

\subsection{Numerical Parameters}

The TB model utilizes computationally and experimentally derived parameters. The SOC parameter is calculated to have the value $\xi = \SI{0.06}{eV}$, which is approximately $\SI{2.2e-3}{\au}$, in~\cite{Aut_s_2006}. The time-dependent simulations begin at $t=0$, end at $t  = T_f = \au{10000}$, and use a timestep of $\delta t = \au{1}$, which is approximately $\SI{24}{as}$.

All other TB parameters are taken from the Oxford model~\cite{OxfordModel}. The chosen TB model was parameterised in a bulk environment for the purpose of reproducing magnetic moments near point defects in solids. The cluster geometry considered in this work uses the same inter-atomic spacing as for bulk Fe. Since the crystal structure of the cluster is not relaxed, the simulations are not expected to be quantitatively accurate. The goal of this work is to investigate the qualitative physics of metallic magnetic clusters in time-varying magnetic fields, and the TB model used is expected to describe this correctly.

\subsection{Computing Observables}

The nuclei in our simulations are treated as classical particles subject to classical forces, but the forces exerted on them by the electrons are evaluated quantum mechanically using the time-dependent equivalent of the Hellman-Feynman theorem~\cite{Hellmann1937,PhysRev.56.340}. Let $\v{R}_a$ denote the position of the nucleus of atom $a$. The Hellman-Feynman theorem states that the force exerted on the nucleus of atom $a$ by the electrons is given by
\begin{align}
	\v{F}_a &= -\textrm{tr}(\rho \v{\nabla}_a H ), \label{eq:force}
\end{align}
where the density matrix is evaluated according to Eq.~\eqref{eq:rho_from_d} and $\v{\nabla}_a = \partial/\partial \v R_a$. The nuclei also experience a classical Lorentz force,
\begin{align}
	\v{F}^{EM}_a = q_a (\v{v}_a\times\v{B}_a) + q_a\v{E}_a,\label{eq:LorentzForce}
\end{align}
where $q_a$ is the charge of nucleus $a$, $\v{B}_a$ and $\v{E}_a$ are the applied magnetic and electric fields at the position of nucleus $a$, and $\v{v}_a$ is the velocity of nucleus $a$.

The classical nuclei experience both an interaction torque, $\v \Gamma_\textrm{int}$, due to the quantum mechanical electrons, and a direct torque, $\v \Gamma^{N,EM}$, exerted by the applied electromagnetic field. 
The total torque acting on the nuclei is the sum of these two contributions:
\begin{align}
	\v \Gamma^N = \v \Gamma^{N,EM} + \v \Gamma_\textrm{int}.
\end{align}
The internal torque is calculated from the Hellman-Feynman forces as
\begin{align}
	\v \Gamma_\textrm{int}(t) =& \sum_{a} \v R_a \times \v F_a(t),
\end{align}
and the direct torque is given by
\begin{align}
	\v{\Gamma}^{N, EM} = \sum_{a \in N} \v{R}_a \cross \v{F}^{EM}_a.\label{eq:directEMTorque}
\end{align}
The angular momentum of the electrons changes as the external field changes, so $\v \Gamma_{\textrm{int}}$ is non-zero.

All other expectation values computed in this work are found by taking the trace of the operator multiplied by the density matrix, for example,
\begin{equation}
	\expval{\v{L}} = \textrm{tr}(\rho \v{L}),\ \expval{\v{S}} = \textrm{tr}(\rho \v{S}),\ \expval{\v{\mu}} = \textrm{tr}(\rho \v{\mu}).
\end{equation}

\subsection{Ehrenfest Equations}

The Ehrenfest equations of motion come in useful when interpreting the simulation results. The algebra required to derive the Ehrenfest equation of motion for the total angular momentum operator, $\v J = \v L + \v S$, is outlined in the appendix of~\cite{twells}. The equations of motion for $\v L$ and $\v S$ separately are derived similarly, although the equation of motion for $\v S$ only receives contributions from the dipole coupling and SOC Hamiltonian terms. The resulting equations are:
\begin{align}
	\frac{d\expval{\v J}}{dt} =& -\v\Gamma_{\textrm{int}} + \expval{\v \mu} \cross \v B, \label{eq:J}\\
	\frac{d\expval{\v L}}{dt}	=& -\v\Gamma_{\textrm{int}} - \frac{\mu_B}{\hbar} \expval{\v L} \cross \v B + \frac{\xi}{i\hbar^3}\expval{[\v L,  \v L \cdot \v S]}, \label{eq:L}\\
	\frac{d\expval{\v S}}{dt} =& - 2\frac{\mu_B}{\hbar}\expval{\v S} \cross \v B + \frac{\xi }{i\hbar^3} \expval{[\v S, \v L \cdot \v S]}. \label{eq:S}
\end{align}

\subsection{Investigative Approach}

A typical period of oscillation in an Einstein-de Haas experiment is of order $\SI{1}{s}$, but well-converged quantum mechanical simulations require a time step of order $\au{1}$ ($\SI{2.4e-17}{s}$). Since simulations of only $100,000$ timesteps ($\num{2.4e-11}$~s) are achievable on consumer hardware in a few hours, the necessary computations might initially seem intractable. Fortunately, however, it is possible to simulate long enough to reach the quasi-adiabatic limit, beyond which further increases in the duration of the simulation do not produce qualitative differences in the results. For example, the total change in angular momentum, calculated by integrating the torque through a 180$^{\circ}$ rotation of the applied magnetic field, becomes independent of the simulation time, which is also the time taken to rotate the field. The instantaneous torque tends to zero as the duration increases, so we cannot work in the fully adiabatic limit and assume that the wave function is the instantaneous ground state at all times, but the simulation results can nevertheless be extrapolated to experimental timescales. To prove that quasi-adiabatic timescales are attainable for the \Fe{15} system, we first characterize the relevant physical timescales.

It is shown in the \hyperref[appendix:adiabatic_theorem]{Appendix} that the timescale associated with precession of the magnetic moment in the applied magnetic field is
\begin{align}
	T_p \sim \frac{2\pi\hbar}{\Delta E_{-\v \mu \cdot \v B}}, \label{eq:T_p}
\end{align}
where $\Delta E_{-\v \mu \cdot \v B}$ is a typical spacing between energy levels of the magnetic dipole Hamiltonian. For states with the same orbital angular momentum quantum number ($m_l$) but different spin quantum numbers ($m_s$), the difference in $m_s$ will always be $\hbar$. In this case, $\Delta E_{-\v \mu \cdot \v B} = 2\mu_B B$, where we have assumed that the spins lie parallel or antiparallel to $\v B$. For an experimentally realistic magnetic field strength of 0.5~T,
\begin{align}
	T_{p} \sim \SI{3.0e6}{\au}
\end{align}
In SI units, this is approximately \SI{7.1e-11}{s}. We note that the adiabatic wave function for the system makes zero contribution to the final angular momentum of the iron, as the start and end points ($B$ aligned along $z$) are equivalent. Thus, the leading term contributing to the net angular momentum transfer will be the first order non-adiabatic correction. The torque applied by the magnetic field on the system must rotate the electron magnetic moments (spin and orbit) with the magnetic field for the beginning and end points to both be adiabatic (Born-Oppenheimer) solutions. No moment is transferred to the lattice by conservation of energy, since there is zero change in adiabatic energy between the beginning and the end, with no kinetic energy being acquired by the nuclei. Thus any gain in kinetic energy of the nuclei must be a result of non-adiabatic processes.
The adiabaticity timescale associated with the electronic structure of the cluster is
\begin{align}
	T_{s} \sim \frac{2\pi \hbar}{\Delta E_{H_0}}, \label{eq:spin-lattice-timescale-formula}
\end{align}
where $\Delta E_{H_0}$, the difference in energy of eigenstates split by $H_0$, takes on energy values in the range \SI{0.1}{\au} to \SI{0.01}{\au} 
Assuming $\Delta E_{H_0} = \SI{0.1}{\au}$ gives
\begin{align}
	T_{s} \sim \SI{62.8}{\au} \label{eq:spin-lattice-timescale}
\end{align}
In SI units, this is approximately \SI{1.5e-15}{s}. 
States of different orbital angular momenta are split by $H_0$ while states with the same spatial form but different spin are not, so this timescale affects states with different values of $\expval{\phi_i|\v L | \phi_i}$, where $\ket{\phi_i}$ is the $i$'th instantaneous eigenstate. The value of $\expval{\v L}$ is able to follow the changes in the applied field quasi-adiabatically, provided the simulation duration is greater than $T_s$. This timescale is much shorter than the timescale associated with Larmor precession.

It is also instructive to calculate the spin-orbit adiabaticity timescale. The eigenvalues of the SOC term are given by  $(\xi/2) (j(j+1) - l(l+1) - s(s+1))$. In our simulations $l = 2$, since we consider only $d$ orbitals, $s = \frac{1}{2}$, and $j$ may take the values of $5/2$ or $3/2$. These two values of $j$ give the SOC eigenvalues $\xi$ and $-1.5 \xi$ respectively. Taking the difference between these gives the energy level separation, $|\Delta E_{SOC}| = 2.5\xi = \SI{0.15}{eV}$, which is the only possible energy level transition coupled by SOC. Using, 
\begin{align}
	T_{SOC} \sim \frac{2\pi \hbar}{\Delta E_{SOC}},
\end{align}
we find that $T_{SOC} = \SI{419.3}{\au}$, which is an order of magnitude larger than the lattice splitting timescale in Eq.~\eqref{eq:spin-lattice-timescale}, but several orders of magnitude smaller than the precession timescale. Thus all simulations that are quasi-adiabatic with respect to the Larmor precession timescale, will also be quasi-adiabatic with respect to the SOC timescale.

Although the precession timescale for a $B$ field of 0.5~T, $\SI{7.1e-11}{s}$, is too long to simulate, it is possible to achieve quasi-adiabaticity in a shorter time by applying an artificially large magnetic field. The largest field strength considered in this work is 500~T, for which the magnetic dipole coupling adiabaticity timescale is $T_p \sim \SI{3e3}{\au}$ This timescale remains greater than the adiabaticity timescales arising from the electronic structure of the crystal ($T_{s} = \SI{62.8}{\au}$), and from the SOC term ($T_{SOC} = \SI{419.3}{\au}$). Since the simulations used to draw any conclusions remain quasi-adiabatic, and the field strengths considered are not sufficiently strong to reorder the adiabaticity timescales, the results of our simulations are qualitatively similar to the results that would be found had an experimentally realistic field strength been used. Our approach will consider a range of $B$ field strengths to confirm the adiabatic timescales calculated above, and deduce trends in the contributions to the torque as the limit of small $B$ and large $T_f$ is approached.

\section{\label{sec:results}Results}

To facilitate a gradual build up in the complexity of the effects observed, the results section is split into two parts: Sec.~\ref{sec:withoutSOC} presents results obtained in the absence of spin-orbit coupling; and Sec.~\ref{sec:withSOC} presents results with spin-orbit coupling. Sec.~\ref{sec:experimental_relevance} examines how the simulations are relevant to experiments. Sec.~\ref{sec:extrapolation} ends the results section with an analysis of the trends of the various contributions to the torque as the experimental limit is approached.

The simulations without SOC used three different field strengths: $B = 500$~T, $B = 50$~T, and $B = 0.5$~T. The results show the gradual breakdown of the quasi-adiabatic rotation of the spin as the applied field is reduced. The simulations with SOC used $B = 500$~T only, as these were unambiguously in the quasi-adiabatic limit and thus the most relevant to experiment. 
Unless stated otherwise, the results below are expressed in Hartree atomic units (a.u.).

\floatsetup[figure]{style=plain,subcapbesideposition=top}
\begin{figure}[t!]
	\centering
    \sidesubfloat[]{\includegraphics[width=0.4\linewidth]{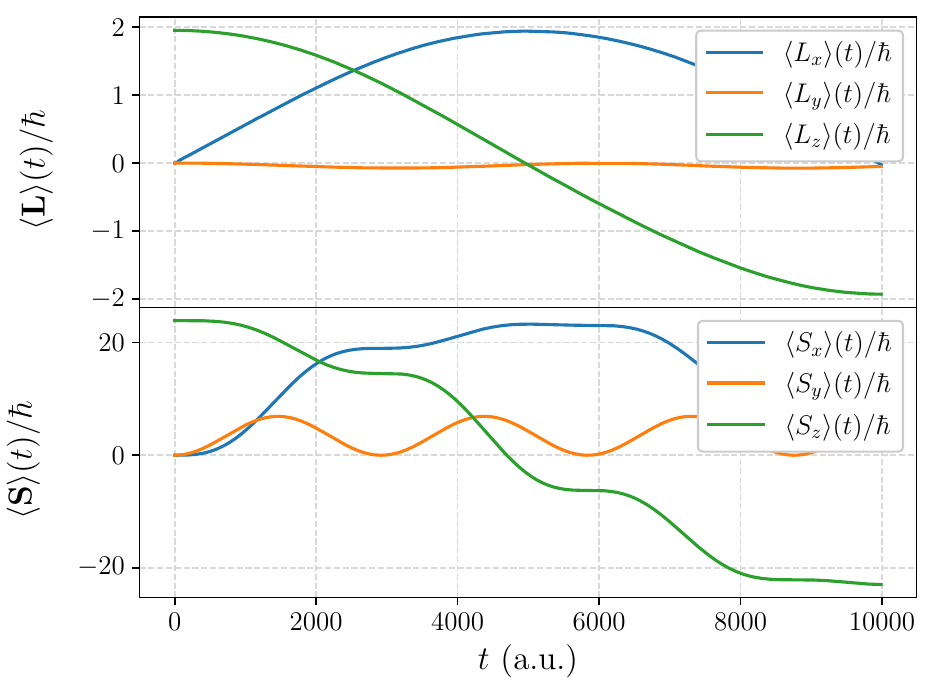}\label{fig:withoutSOC_B=500_a}}
    \quad
    \sidesubfloat[]{\includegraphics[width=0.4\linewidth]{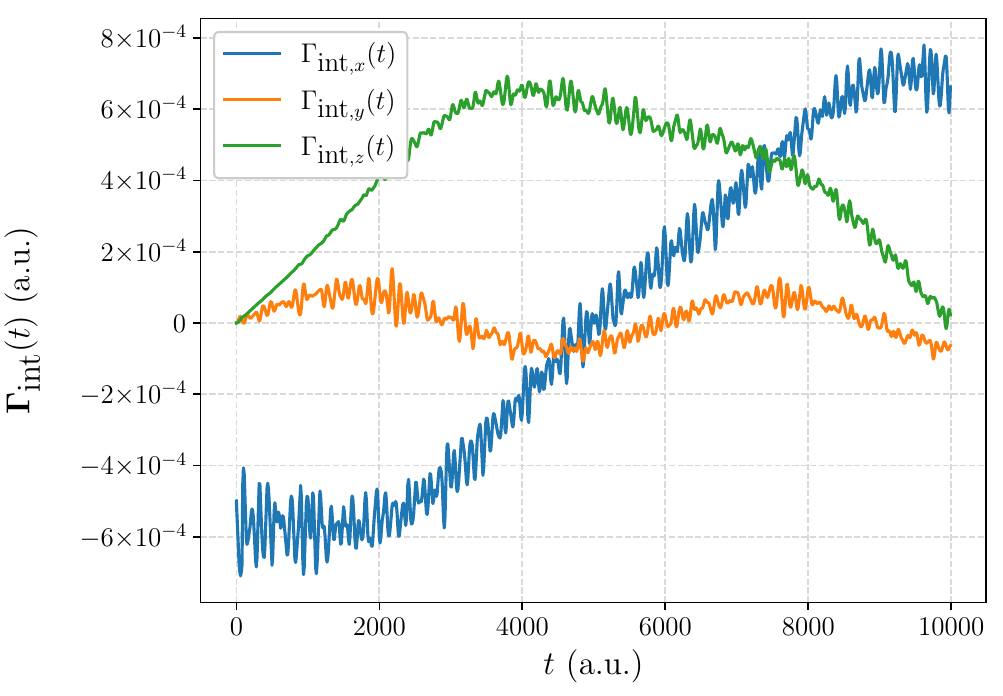}\label{fig:withoutSOC_B=500_b}}
	\caption{The time evolution of the expectation values of (a) the  orbital and spin angular momenta and (b) the torque as the applied magnetic field rotates in the $xz$ plane at constant angular velocity. The field strength is $500$~T and there is no SOC. The orbital angular momentum remains almost perfectly anti-aligned with $\v B$. The spin is approximately anti-aligned but exhibits additional oscillations due to Larmor precession about the $\v B$ field. The simulation averages of the $x$ and $y$ components of the torque exerted on the nuclei by the electrons are approximately $0$; the average of the $z$ component is non-zero.} 	\label{fig:withoutSOC_B=500}
\end{figure}

\subsection{Without spin-orbit coupling}\label{sec:withoutSOC}

The results shown in this section were all obtained in the absence of SOC, i.e., with $\xi = \SI{0}{\au}$ The effects of exchange and the interaction with the magnetic field were included. The three simulations considered have (i) $B = 500$~T, (ii) $B = 50$~T, and (iii) $B = 0.5$~T.

Figure~\ref{fig:withoutSOC_B=500_a} shows the evolution of the spin and orbital angular momentum expectation values in response to a time-varying $B$ field with a field strength of $B = 500$~T. Although $\expval{\v S}$ remains approximately antiparallel to the field, it also oscillates slightly with a period of approximately $\num{3000}$~a.u. This is the timescale associated with the Larmor precession of the spins: the Larmor frequency, $\omega_S = 2 \mu_B B/\hbar$, implies a period of oscillation of $\frac{2\pi\hbar}{2\mu_B B} = \au{2954} \approx \SI{7.1e-14}{s}$ 

From Eqs.~\eqref{eq:L} and~\eqref{eq:S}, setting the spin orbit term to zero, one can see that $\expval{\v S}$ can only undergo Larmor precession, whereas $d\expval{\v L}/dt$ has contributions from a Larmor term plus the interaction torque. The interaction torque is much larger than the magnetic torque, explaining why $\expval{\v L}$ does not precess at the Larmor frequency.

The torque exerted on the nuclei by the electrons, $\v \Gamma_{\textrm{int}}$, is shown in figure~\ref{fig:withoutSOC_B=500_b}. The  $y$ component remains small throughout the simulation; the $x$ component changes from negative to positive as the field rotates; and the time dependence of the $z$ component is shaped (approximately) like the first half of a sinusoidal cycle. 
If we were not holding the atoms in place, and if the velocities of the nuclei remained small enough to justify neglect of the direct Lorentz torque, the ``torque impulse'' $\Delta \v{L}_{\textrm{nuclei}} = \int_0^{T_{\textrm{f}}} \v\Gamma_{\textrm{int}}(t) dt$ would equal the change in the angular momentum of the cluster of classical nuclei during the simulation. The contributions from $\Gamma_{\textrm{int},x}$ and $\Gamma_{\textrm{int},y}$ are much smaller than the contribution from $\Gamma_{\textrm{int},z}$ and integrate to zero in the quasi-adiabatic limit, so the cluster would begin to spin about the $z$ axis.

In addition to the torque on the nuclei due to the electrons, the nuclei also experience a direct torque contribution from the EM field via the Lorenz force.  The effect of the direct electromagnetic torque can be estimated from Eqs.~\eqref{eq:LorentzForce} and~\eqref{eq:directEMTorque}. Since the nuclei are clamped, then $\v v_a = \v 0$ for all atoms and thus the classical Lorentz force is given by $\v{F}^{EM}_a = q_a\v{E}_a$. Faraday's law of induction informs us that $\v\nabla \cross \v E = -\frac{\partial \v B}{\partial t}$. Since the $B$ field is spatially uniform, it follows that $\frac{\partial \v B}{\partial t}$ is spatially uniform, thus the curl operator can be inverted to give $\v E = -\frac{1}{2}(\frac{\partial \v B}{\partial t})\times \v r + \v{\nabla}\chi(\v r,t)$, where $\chi(\v r,t)$ is an arbitrary smooth function of $\v r$ and $t$. Since there are no charges contributing to the external field in the vicinity of the cluster, we require the solution with ${\v \nabla}\cdot{\v E} = 0$, which sets $\chi(\v r,t)=0$ if the boundary condition that the electric field should tend to zero as $r$ becomes large is also applied. These relations can be used to estimate the direct torque on the nuclei due to the EM field. Using Eq.~\eqref{eq:directEMTorque} for the torque on the nuclei, we find
\begin{align}
    \v\Gamma^{N,EM} =& -\frac{1}{2}\sum_{a\in N} q_a \v R_a \cross \bigg(\frac{\partial \v B_a}{\partial t}\times \v R_a\bigg),
\end{align}
which has a magnitude of order
\begin{align}
    \Gamma^{N,EM} \sim& eN R_c^2 \bigg|\frac{\partial \v B}{\partial t}\bigg|,
\end{align}
where $N$ is the number of nuclei and $R_c$ is the mean cluster radius. A field of $\SI{0.5}{T}$ that reverses its direction over a duration of $\SI{10000}{\au}$, has $\frac{\partial \v B}{\partial t} \sim \SI{4.3e-11}{\au}$ An approximate mean cluster radius of $\SI{2.49}{\angstrom}$, gives $\Gamma^{N,EM} \sim \SI{1.8e-9}{\au}$ In figure~\ref{fig:withoutSOC_B=0.5}, which has a field strength of $B = \SI{0.5}{T}$, the interaction torque is approximately $\SI{0.5e-6}{\au}$ on average, thus the contribution of the direct EM torque is negligible in comparison to the interaction torque due to the electrons. Provided that simulations remain quasi-adiabatic, both the interaction torque and Faraday torque scale as $1/T_f$, so this result also holds for experimental timescales.


Since $\expval{\v L}$ is mostly antiparallel to $\v B$, the $-\frac{\mu_B}{\hbar}  \expval{\v L}\cross \v B$ precession term in Eq.~\eqref{eq:L} is small and $\v\Gamma_{\textrm{int}}$ is approximately equal to $-\frac{d\expval{\v L}}{dt}$. This can be seen by comparing the graphs of $\expval{L_x}(t)$ and $\Gamma_{\textrm{int},x}(t)$: since $\expval{L_x}(t)$ is shaped like the first half of a sine curve, $\Gamma_{\textrm{int},x}(t)$ is shaped like minus the first half of a cosine curve. Similarly, $\expval{L_z}(t)$ is shaped like the first half of a cosine curve and $\Gamma_{\textrm{int},z}(t)$ like the first half of a sine curve. These observations show that the internal torque on the nuclei is a result of the transfer of orbital angular momentum from the electrons to the nuclei. The transfer could be seen as a manifestation of the Einstein-de Haas effect, but for orbital angular momentum rather than spin. It arises from the rotation of the orbital magnetic moment created by the application of the field itself, and is transmitted to the nuclei via the Coulomb interactions between electrons and nuclei. In the absence of spin-orbit interactions, although the spins rotate, they are decoupled from the lattice and do not exert torques on the nuclei. The torque exerted by the rotating applied field changes the spin angular momentum directly, with no involvement of the lattice.

The rapid oscillations appearing in the torque do not arise from the $-d\expval{\v L}/dt$ term of Eq.~\eqref{eq:L}, which does not vary on this timescale, but from the small precession term, $\frac{1}{2}\v B \times \expval{\v L}$. Their existence indicates that as $\v B(t)$ evolves, $\expval{\v L}(t)$ is not perfectly anti-parallel to $\v B(t)$, but remains approximately anti-parallel by continuously correcting itself on the crystal Hamiltonian timescale (\SI{1.5e-15}{s}, or about 62.8 a.u.), which was calculated in Eq.~\eqref{eq:spin-lattice-timescale-formula}.

\floatsetup[figure]{style=plain,subcapbesideposition=top}
\begin{figure}[t!]
	\centering
    \sidesubfloat[]{\includegraphics[width=0.4\linewidth]{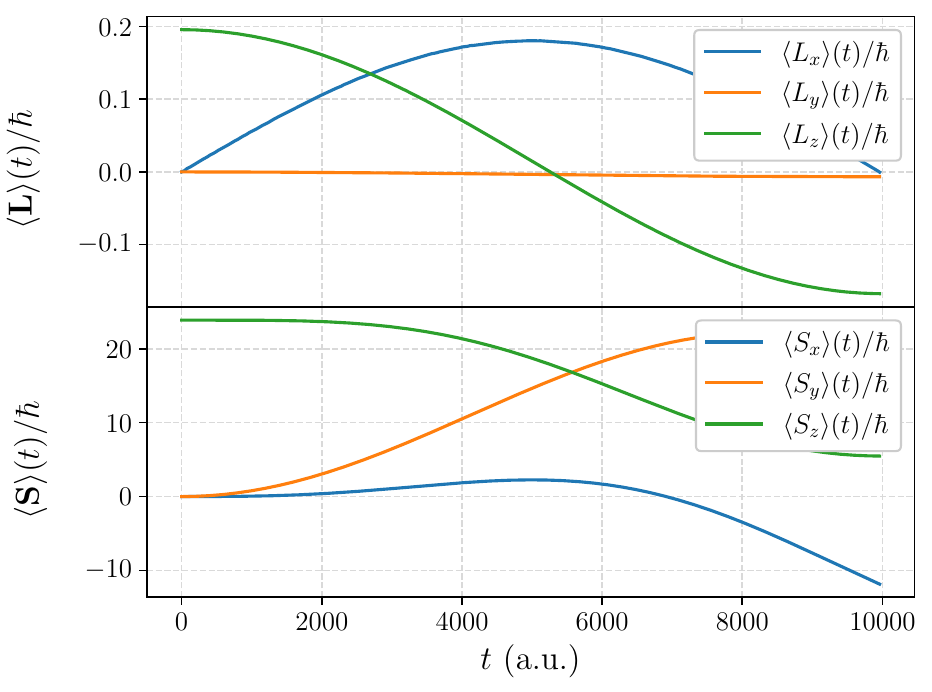}\label{fig:withoutSOC_B=50_1}}
    \quad
    \sidesubfloat[]{\includegraphics[width=0.4\linewidth]{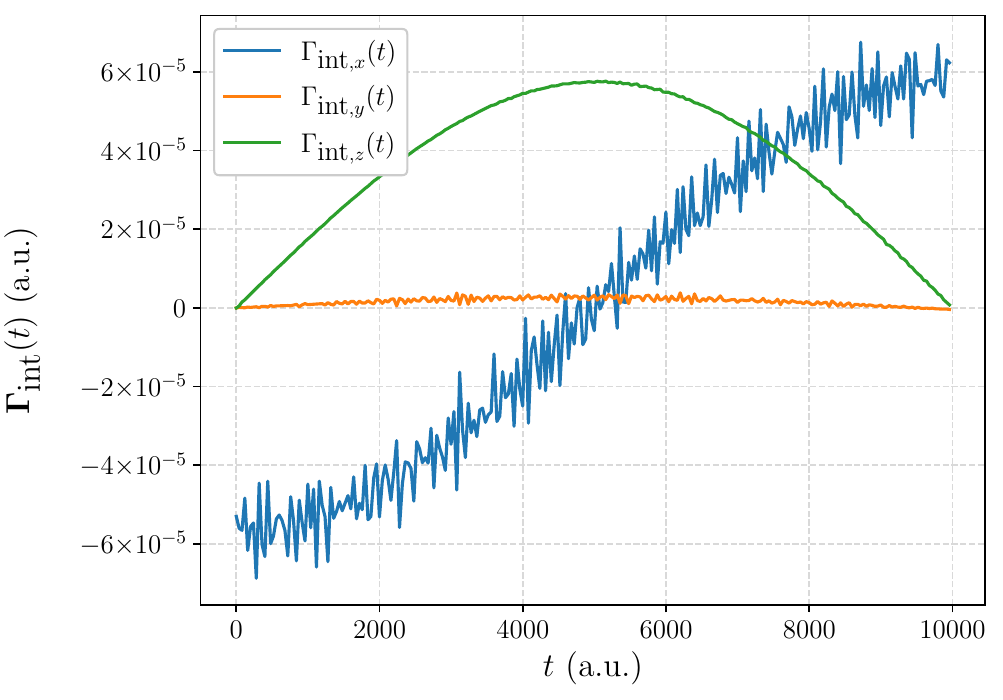}\label{fig:withoutSOC_B=50_2}}
	\caption{The time evolution of the expectation values of (a) the orbital and spin angular momenta and (b) the torque as the applied magnetic field rotates in the $xz$ plane at constant angular velocity. The field strength is B = 50~T and there is no SOC. The orbital angular momentum again remains approximately anti-aligned with $\v B$. The spin fails to stay anti-aligned with $\v B$ as the Larmor precession is too slow for this field strength and the simulation is not quasi-adiabatic. The simulation averages of the $x$ and $y$ components of the torque are approximately 0; the average of the $z$ component is non-zero.}
	\label{fig:withoutSOC_B=50}
\end{figure}

The results of the $B = \SI{50}{T}$ simulation are shown in figure~\ref{fig:withoutSOC_B=50}. The spin precession timescale of $\frac{2\pi\hbar}{2\mu_B B} = \au{29537}$ (${\SI{7.1e-13}{s}}$) is 10 times greater than it is when $B = \SI{500}{T}$, and is almost half the duration of the simulation.
The spin is unable to keep up with the rotating magnetic field, and the simulation ends without the $z$ component of the spin reversing its sign. Since the spin fails to stay in its ground state, the behaviour of the spin is not quasi-adiabatic at this magnetic field strength and rate of change. The initial magnitude of the spin is determined mostly by the exchange interaction and remains similar to the $B = \SI{500}{T}$ case, but the orbital angular momentum $\expval{\v L}$ is reduced by a factor of 10. This explains the reduction by about a factor of 10 in the torque applied to the nuclei. The difference in the evolution of the spin has little qualitative effect on the evolution of $\expval{\v L}$ and thus little qualitative effect on the form of the torque.

\floatsetup[figure]{style=plain,subcapbesideposition=top}
\begin{figure}[t!]
	\centering
    \sidesubfloat[]{\includegraphics[width=0.4\linewidth]{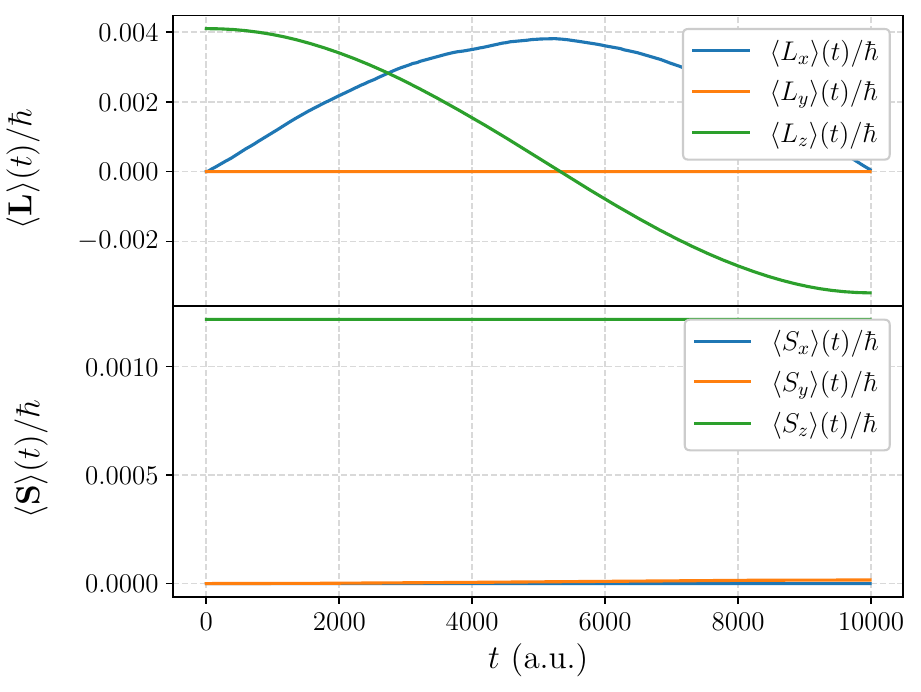}\label{fig:withoutSOC_B=0.5_1}}
    \quad
    \sidesubfloat[]{\includegraphics[width=0.4\linewidth]{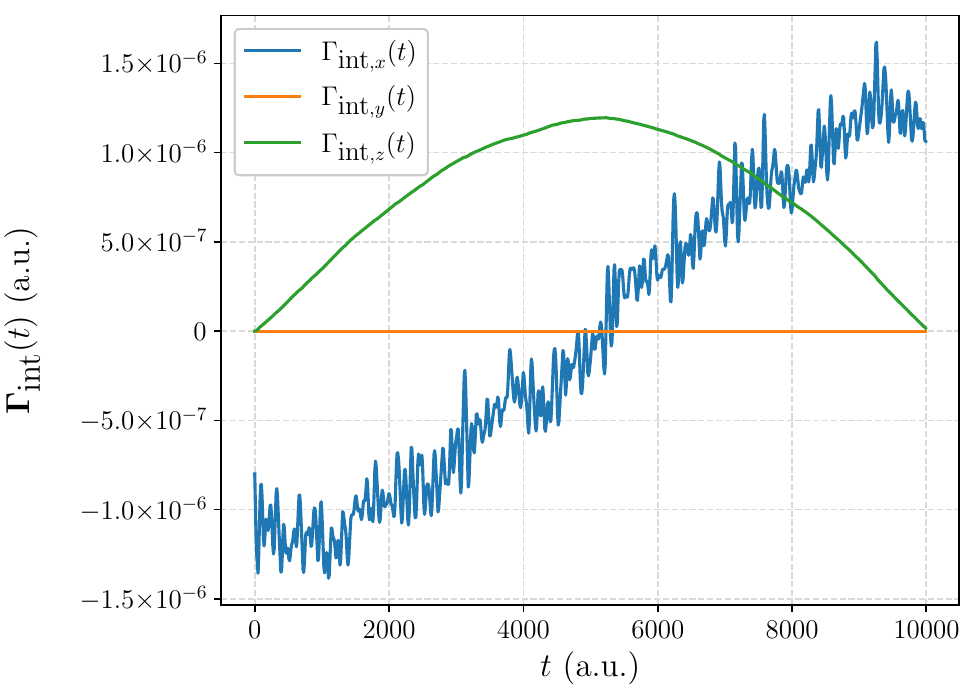}\label{fig:withoutSOC_B=0.5_2}}
	\caption{The time evolution of the expectation values of (a) the orbital and spin angular momenta and (b) the torque as the applied magnetic field rotates in the $xz$ plane at constant angular velocity. The field strength is $B = 0.5$~T and there is no SOC. The orbital angular momentum remains approximately anti-aligned with $\v B$. The spin is almost completely unable to respond to the rotation of $\v B$ as the Larmor precession period is greater than the simulation duration. The simulation averages of the $x$ and $y$ components of the torque are approximately 0; the average of the $z$ component is non-zero.} 
	\label{fig:withoutSOC_B=0.5}
\end{figure}

Figure~\ref{fig:withoutSOC_B=0.5} shows the results for a realistic field strength of ${B = \SI{0.5}{T}}$, although still an unrealistically fast field rotation rate. In this case the precession timescale is $\frac{2\pi\hbar}{2\mu_B B} = \au{295375}$ (${\SI{7.1e-11}{s}}$), which is greater than the duration of the simulation. As a result, the evolution of $\expval{\v S}$ is far from adiabatic and the direction of the spin is unable to follow the rotation of $\v B$. The electronic structure timescale (of approximately $\au{62.8}$) is still much smaller than the timescale on which the $B$ field rotates, so the orbital angular momentum $\expval{\v L}$ is able to stay anti-parallel to $\v B$.

In most real solids, the orbital angular momentum is quenched and $\expval{\v L}$ is approximately zero in the absence of an applied magnetic field. Applying a $B$ field induces an $L$, which is proportional to $B$ in the linear regime. The proportionality of $L$ and $B$ can be seen in our Fe$_{15}$ cluster results when $B \gtrapprox \SI{50}{T}$, although $L$ becomes larger than expected when $B$ is small, presumably because the outermost shell of degenerate states is partially filled and can be occupied by electrons in a manner that produces a finite but small orbital angular momentum at very little cost in energy.

For the relatively low magnetic field strengths accessible experimentally, the induced orbital angular momentum is small and the orbital EdH effect discussed in this section is weak. The spin angular momentum, by contrast, is non-zero even in the absence of an applied magnetic field because of the exchange interaction. In the presence of SOC, the rotation of the spin moment also applies a torque to the lattice and produces the spin EdH effect discussed below.

\subsection{With spin-orbit coupling}\label{sec:withSOC}

\floatsetup[figure]{style=plain,subcapbesideposition=top}
\begin{figure}[t!]
	\centering
    \sidesubfloat[]{\includegraphics[width=0.4\linewidth]{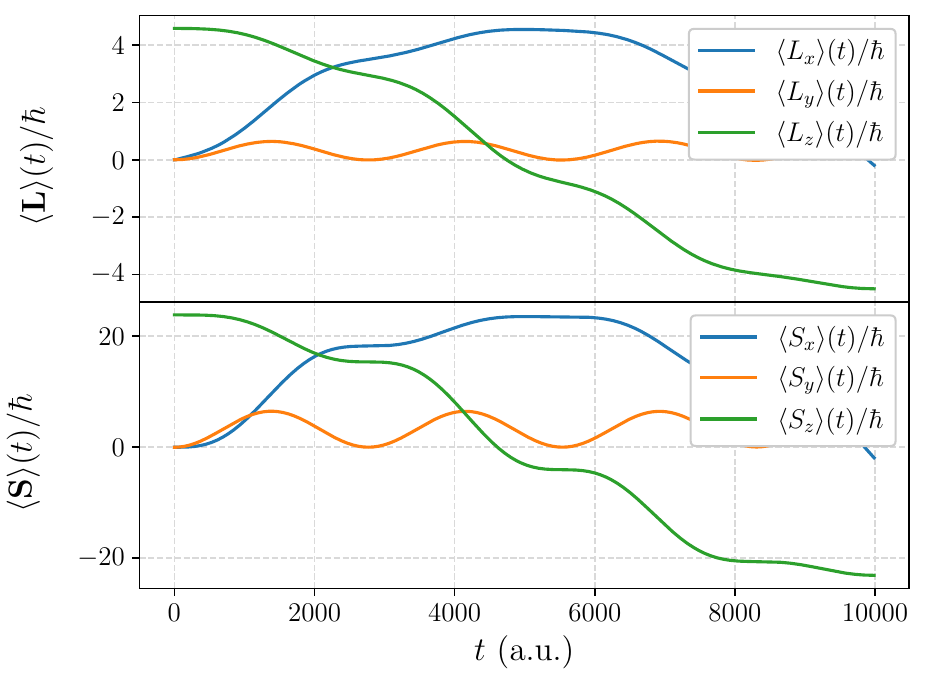}\label{fig:withSOC_B=500_1}}
    \qquad
    \sidesubfloat[]{\includegraphics[width=0.4\linewidth]{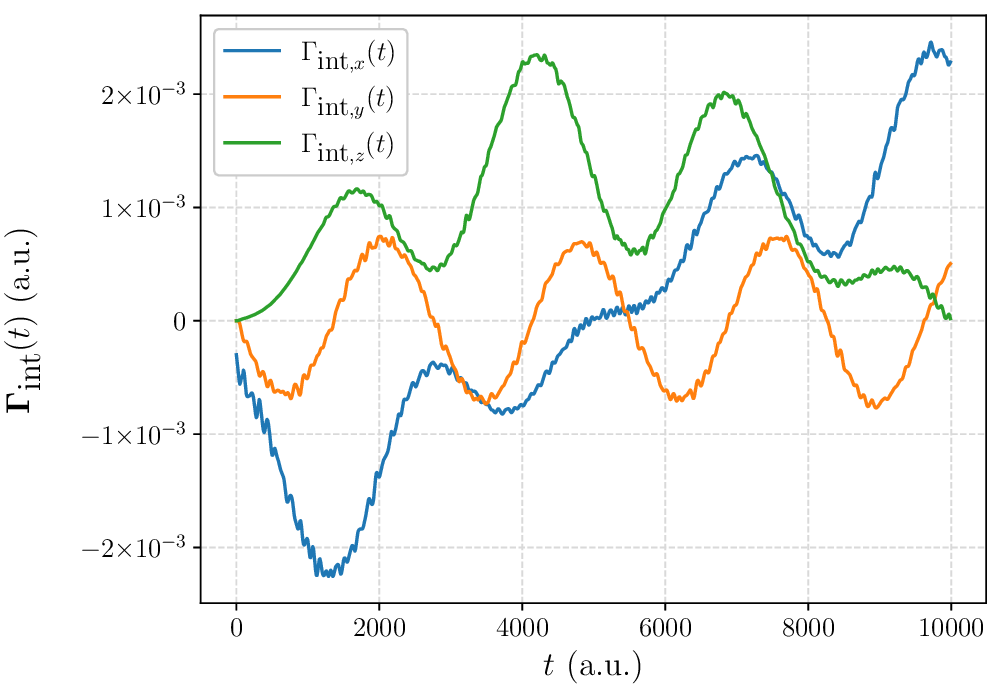}\label{fig:withSOC_B=500_2}}
	\caption{The time evolution of the expectation values of (a) the orbital and spin angular momenta and (b) the torque as the applied magnetic field rotates in the $xz$ plane at constant angular velocity. The field strength is $B = 500$~T and the simulation includes the effects of SOC. The orbital angular momentum is larger than in the absence of SOC and experiences additional oscillations due to its coupling to $\v S$. The effect of Larmor precession about the $B$ field is visible in the evolution of the torque. The simulation averages of the $x$ and $y$ components of the torque exerted on the nuclei by the electrons are approximately 0; the average of the $z$ component is non-zero.}
	\label{fig:withSOC_B=500}
\end{figure}

The results in this section include the effects of SOC, with the SOC parameter $\xi = \SI{0.06}{eV}$ (approximately ${\SI{2.2e-3}{\au}}$) as is appropriate for iron. Figure~\ref{fig:withSOC_B=500} shows the results of a simulation with a $B$ field of 500~T. The spin evolves similarly to the corresponding simulation without SOC (figure~\ref{fig:withoutSOC_B=500_a}). The coupling of $\v L$ and $\v S$ has two main effects. The first is that the initial magnitude of $\expval{\v L}$ is over twice as large as in the equivalent simulation without SOC. This is because the $\v L$ operator not only has the $\v B$ field acting on it, but is also coupled to the $\v S$ operator, the expectation value of which is large because the exchange interaction is large. 

We note that a classical spin-orbit term, of the form $\frac{\xi}{\hbar^2} {\v L \cdot \v S}$, would encourage $\v L$ and $\v S$ to anti-align (for $\xi > 0$). However, in our results, the addition of spin-orbit causes the angular momenta $\v L$ and $\v S$ to couple more strongly in alignment with each other. This can be understood as being due to Hund’s third rule, which states that the value of $J$ is found using $J = \lvert L - S \rvert$ if the shell is less than half full and  $J = \lvert L + S \rvert$ if the shell is more than half full~\cite{BlundellMagnetism}. The \Fe{15} cluster considered has 102 electrons occupying the 150 available MOs, thus the shell is over half full, and thus the energy is minimized with $\expval{\v L}$ and $\expval{\v S}$ in alignment. We were able to verify that Hund’s third rule is obeyed by our model in the case in which the shell is less than half full by additional calculations involving fewer than 75 electrons. In these simulations, the addition of SOC caused $\v L$ and $\v S$ to become anti-aligned, in agreement with Hund’s third rule and as would be expected from the classical interpretation of the SOC term.

The second effect caused by the addition of SOC is that the oscillations due to the Larmor precession about the $B$ field, are also visible in the evolution of $\expval{\v L}$. In figure~\ref{fig:withoutSOC_B=500_a}, the effect of Larmor precession was apparent in the evolution of the spin, yet, the same oscillations did not appear in the evolution of the orbital angular momentum, since the interaction torque is much larger than the magnetic torque in Eq.~\eqref{eq:L}. When SOC is included, the SOC term in Eq.~\eqref{eq:L} becomes significant, and it is energetically favourable for the orbital angular momentum to remain aligned with the spin, which causes the Larmor precession oscillations to also show in the evolution of $\expval{\v L}$.

In the presence of spin-orbit coupling, Eq.~\eqref{eq:L} gives
\begin{align}
	\v\Gamma_{\textrm{int}} &= -\frac{d\expval{\v L}}{dt} - \frac{\mu_B}{\hbar}  \expval{\v L}\cross \v B + \frac{\xi}{i\hbar^3}\expval{[\v L,  \v L \cdot \v S]}. \label{eq:torque4}
\end{align}
As a result, the Larmor oscillations in $\expval{\v L}$ also influence the torque. When averaged over the duration of the simulation, the introduction of SOC more than doubles the magnitude of the interaction torque, as may be seen by comparing figures~\ref{fig:withoutSOC_B=500_b} and \ref{fig:withSOC_B=500_2}. The precession and SOC terms act in opposite directions and mostly cancel each other out, so the interaction torque remains approximately equal to the  $-\frac{d\expval{\v L}}{dt}$ term in Eq.~\eqref{eq:torque4}. 

As explained in Sec.~\ref{sec:withoutSOC}, $\expval{\v L}$ is approximately proportional to $\v B$ when SOC is omitted and $B$ is small. The spin expectation value, by contrast, is determined primarily by exchange interactions and remains substantial even at $B = 0$. Adding SOC links the spin and orbital angular momenta, allowing the spin to mimic an applied field that polarizes the orbital angular momentum and makes the magnitude of the orbital angular momentum independent of $B$ at low $B$. 


\subsection{\label{sec:experimental_relevance}Relevance to experiments}

Although adiabatic simulations at realistic field strengths are impractical, our quasi-adiabatic results allow us to deduce the main qualitative features of the Einstein-de Haas effect on experimental timescales and for experimental field strengths. An experiment with a field strength of $B=\SI{0.5}{T}$ would have a precession timescale of $\num{47014}~\textrm{a.u.}$, much less than the period of the oscillatory fields used in experiments, which are typically of order $\SI{1}{s} = \au{4.1e16}$ It follows that the experimental time evolution is also quasi-adiabatic and that our quasi-adiabatic simulations access the same physics as the experiment. The spin and orbital angular momentum are both able to follow the rotation of the field and reverse their orientations as the field reverses. This generates a measurable torque on the \Fe{15} nuclei. If the \Fe{15} cluster were not held in place, this torque would cause it to start rotating.

\subsection{\label{sec:extrapolation}Extrapolating the results to low $B$ field}

Having established that we are able to carry out simulations in the physically relevant quasi-adiabatic regime, we investigate the trends as the magnitude of $B$ reduces towards $\SI{1}{T}$ or lower, as used in most experiments. 

For a sufficiently small $B$ field, the simulations fail to remain quasi-adiabatic and the results are no longer relevant to experiment. If we suppose that the quasi-adiabatic breakdown occurs when the simulation duration is smaller than the precession timescale, Eq.~\eqref{eq:T_p} tells us that breakdown should occur when ${B \lessapprox \frac{2\pi\hbar}{2 \mu_B T_f } = \SI{9.8}{T}}$. For safety's sake, it is best to ignore results calculated with values of $B$ less than around 20 T.

In figures~\ref{fig:fig_5} and \ref{fig:fig_6}, we plot the results of simulations for a wide range of magnetic field strengths from ${B = \SI{250}{T}}$ to ${B = \SI{0}{T}}$, at a fixed simulation duration of ${T_f = \au{150000}}$
For every field strength considered and every simulation, we calculate the simulation averages of all terms appearing on the right-hand sides of the Ehrenfest equations of motion for $d\expval{\v J}/dt$, $d\expval{\v L}/dt$, and $d\expval{\v S}/dt$, Eqs.~\eqref{eq:J}\textendash\eqref{eq:S}. Every such term may be interpreted as a torque. Only the $z$ components are required, as the time-averaged $x$ and $y$ components are approximately zero. For an initial time of $t=\au{0}$, and a final simulation time of $T_f$, the simulation average is defined by
\begin{align}
	\mean{\Gamma}_z = \frac{1}{T_f} \int_0^{T_f} \Gamma_z(t') dt',
\end{align}
where $\Gamma_z(t')$ is a time-dependent torque contribution.

For simplicity, just as in Secs.~\ref{sec:withoutSOC} and \ref{sec:withSOC}, the results obtained without SOC will be described before the results with SOC.


\subsubsection{\label{sec:low_field_without_SOC}Without spin-orbit coupling}

\begin{figure}[t!]
    \centering
    \sidesubfloat[]{\includegraphics[width=0.5\linewidth]{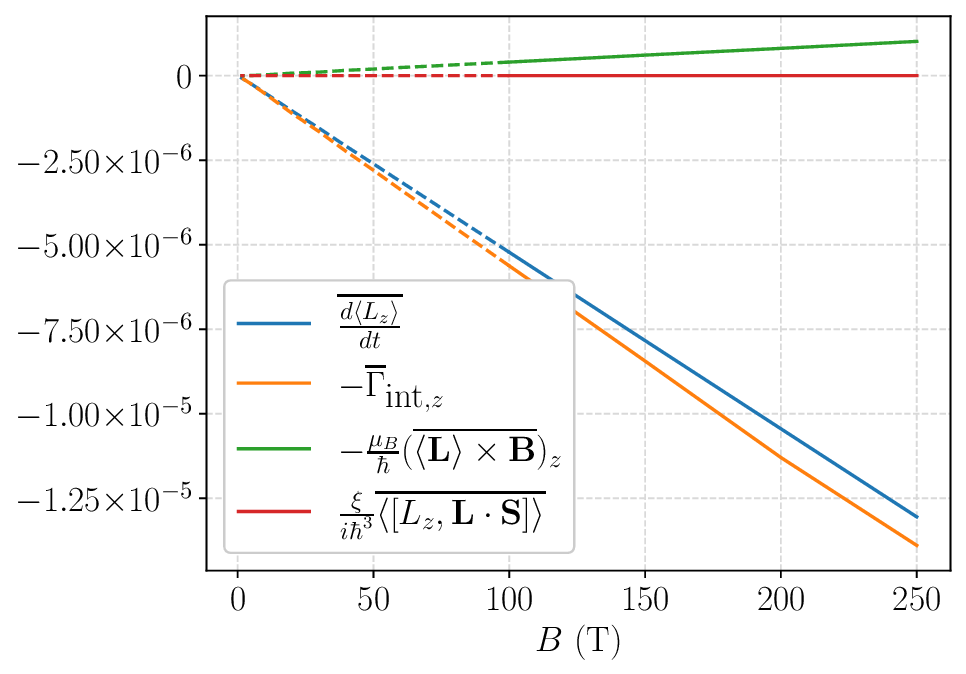}\label{fig:fig_5a}}
    \quad
    \sidesubfloat[]{\includegraphics[width=0.5\linewidth]{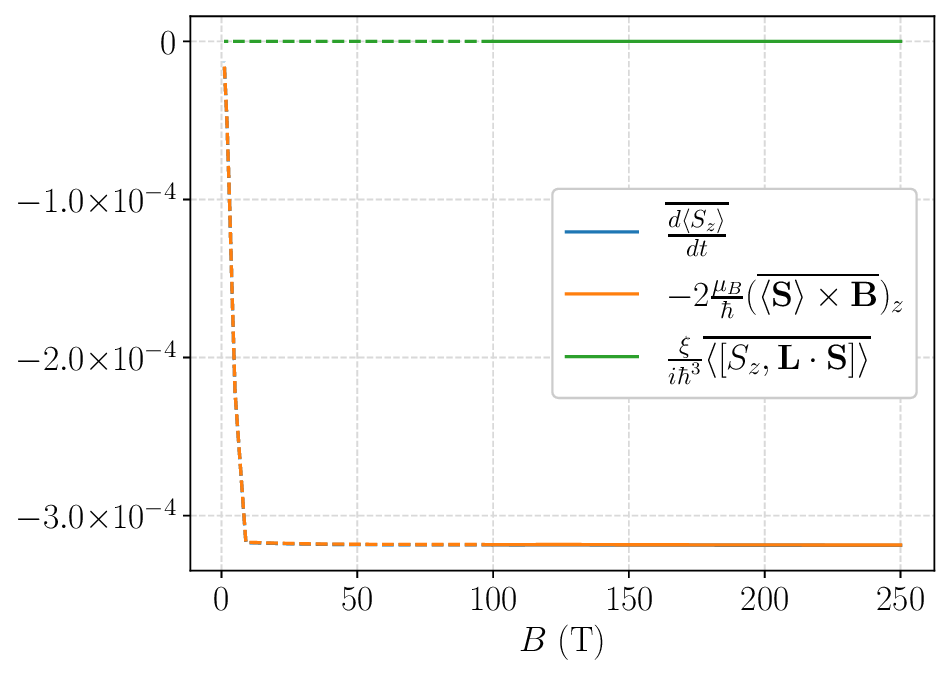}\label{fig:fig_5b}}
    \vskip\baselineskip
    \sidesubfloat[]{\includegraphics[width=0.5\linewidth]{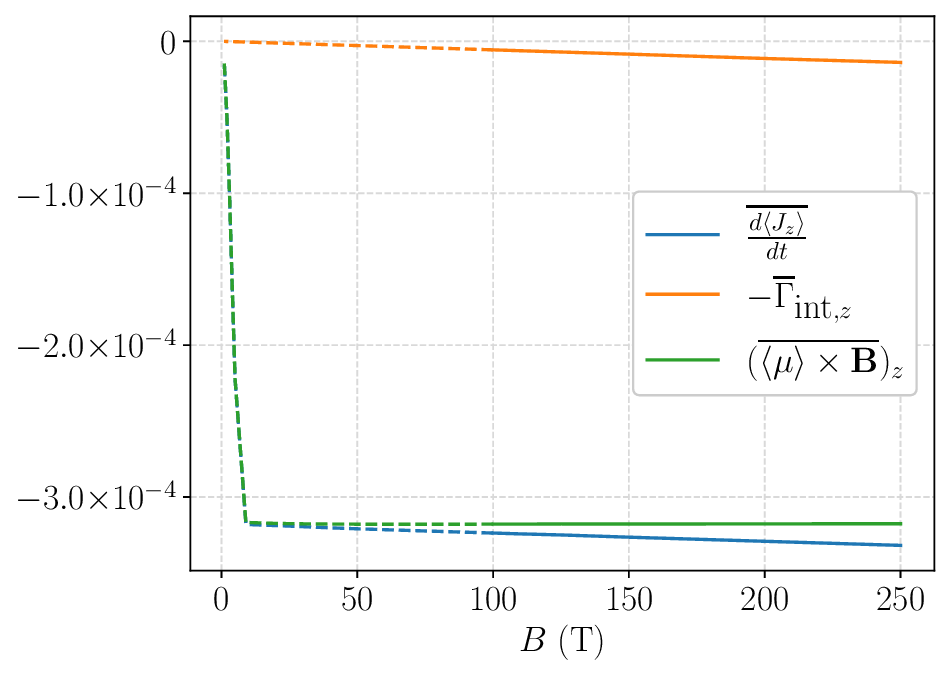}\label{fig:fig_5c}}
	\caption{The time-averaged torques entering the equations of motions for (a) $\expval{\v L}$, (b) $\expval{\v S}$ and (c) $\expval{\v J}$, 
		for a range of $B$ field strengths and with a fixed simulation duration of ${T_f = \au{150000}}$ The results in this figure are without SOC. Torque values for $B<\SI{100}{T}$ are shown with dashed lines to indicate that the data in this region should not be analysed.}\label{fig:fig_5}
\end{figure}

\begin{figure}[t!]
	\centering
    \sidesubfloat[]{\includegraphics[width=0.5\linewidth]{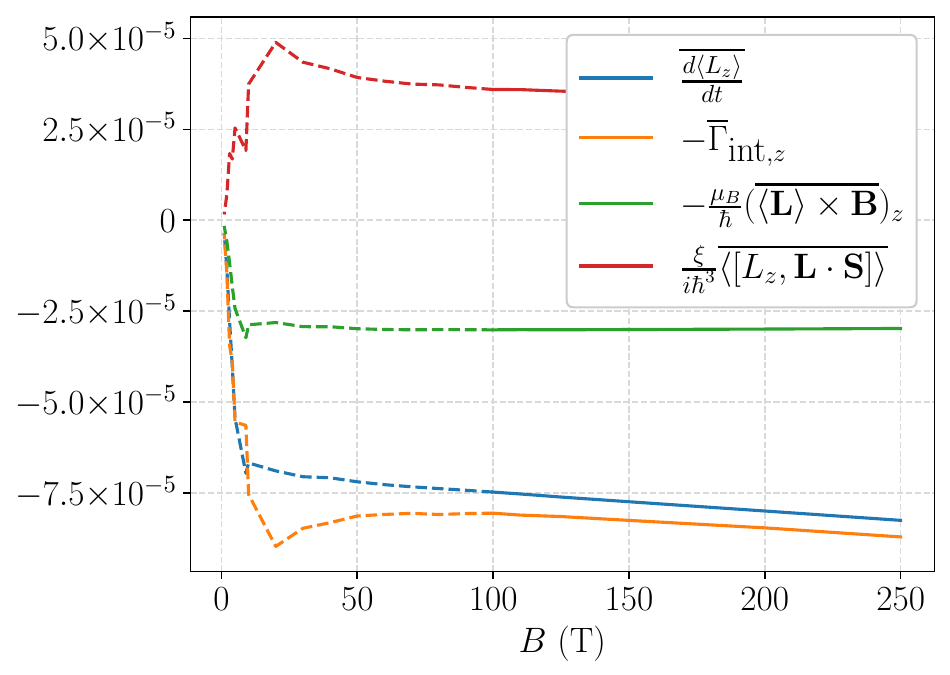}\label{fig:fig_6a}}
    \quad
    \sidesubfloat[]{\includegraphics[width=0.5\linewidth]{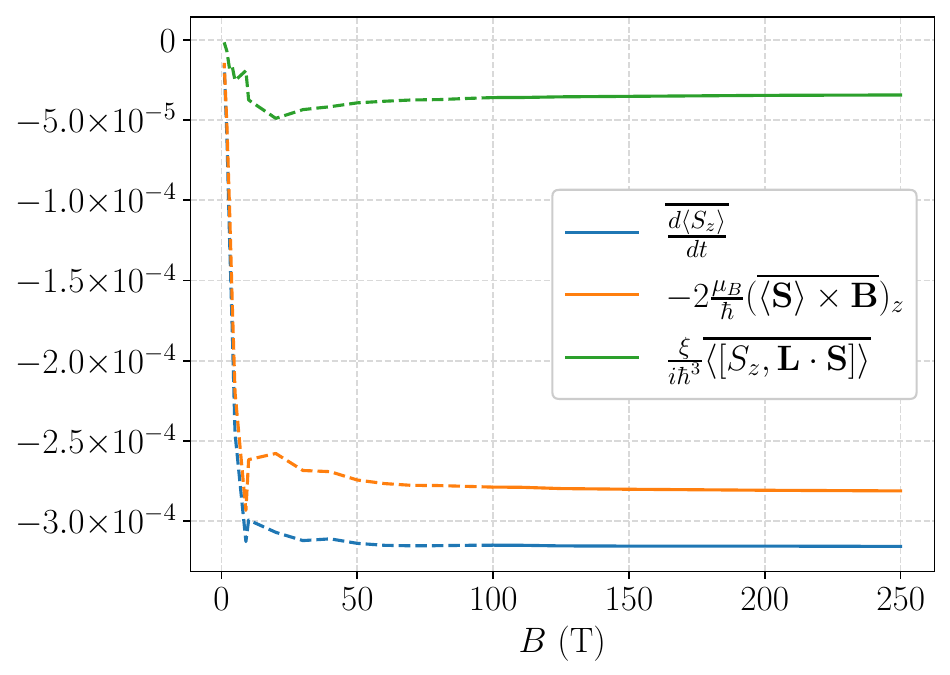}\label{fig:fig_6b}}
	\vskip\baselineskip
    \sidesubfloat[]{\includegraphics[width=0.5\linewidth]{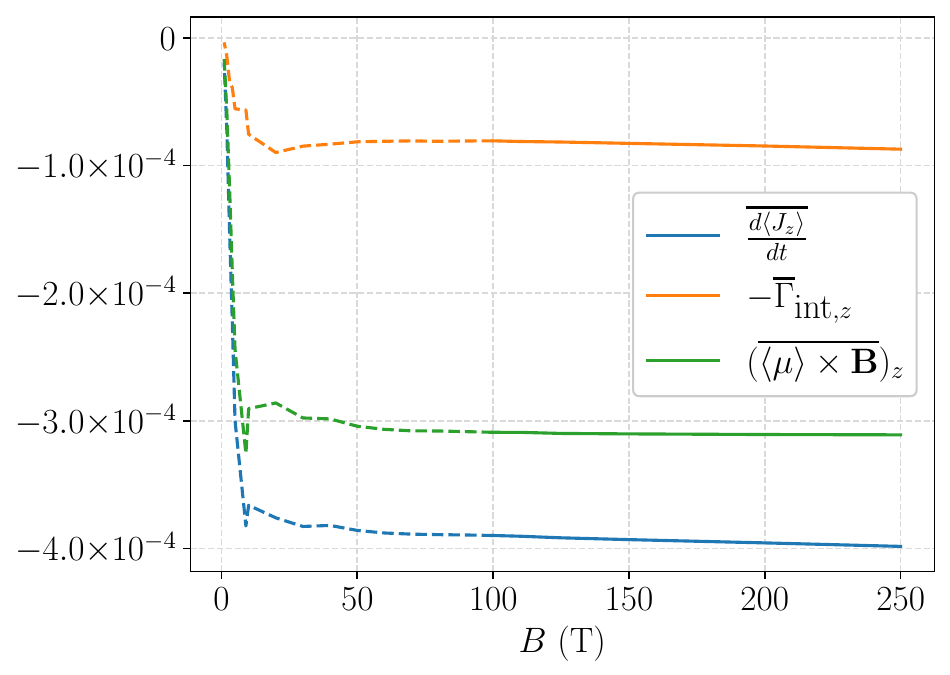}\label{fig:fig_6c}}
	\caption{The time-averaged torques entering the equations of motions for (a) $\expval{\v L}$, (b) $\expval{\v S}$, (c) $\expval{\v J}$,
		for a range of $B$ field strengths and with a fixed simulation duration of ${T_f = \au{150000}}$ The results in this figure are with SOC. Torque values for $B<\SI{100}{T}$ are shown with dashed lines to indicate that the data in this region should not be analysed.}\label{fig:fig_6}
\end{figure}

Figure~\ref{fig:fig_5a} shows how the simulation-averaged contributions to $d\expval{\v L}/dt$ given in Eq.~\eqref{eq:L} depend on the magnitude of $B$ in the absence of SOC.
%
The $z$ component of the term that arises from the coupling of the orbital dipole to the applied magnetic field is given by ${(-\mu_B/\hbar)(\expval{\v L}\times \v B)_z} = {(-\mu_B/\hbar)(\expval{L_x} B_y - B_x \expval{L_y})}$. Since $B_y = 0$ and $B_x < 0$ throughout the motion, and since $\expval{L_y} < 0$ (see figure~\ref{fig:withoutSOC_B=500_a}), the resulting torque points in the $+\u z$ direction. 
The dipole torque is small in magnitude because, in a quasi-adiabatic simulation, $\expval{\v L}$ and $\v B$ remain almost anti-parallel and their cross product is small.
The averaged interaction torque applied to the electrons by the nuclei, $-\mean\Gamma_{\textrm{int},z}$, acts in the opposite direction to the magnetic dipole torque, $- \frac{\mu_B}{\hbar}(\mean{\expval{\v L}\cross \v B})_z$.
In the absence of SOC, all the torques in figure~\ref{fig:fig_5a} scale linearly with $B$. $\expval{\v L}$ is small for low $B$ fields, so very little torque is generated by the electrons on the nuclei for experimentally realistic magnetic field strengths.

\floatsetup[figure]{style=plain,subcapbesideposition=top}

\floatsetup[figure]{style=plain,subcapbesideposition=top}

Figure~\ref{fig:fig_5b} shows the torque contributions affecting the spin in Eq.~\eqref{eq:S}.
Since the SOC parameter $\xi$ is zero, the other two terms, ${\mean{\frac{d\expval{S_z}}{dt}}}$ and ${- 2\frac{\mu_B}{\hbar}(\mean{\expval{\v S}\cross \v B})_z}$ are equal. Thus, the rotation of $\expval{\v S}$ is caused solely by the magnetic dipole coupling of the spin to the field. 
The averaged torque due to the dipole moment coupling to the magnetic field acts in the $-\u z$ direction. This is the opposite sign to the dipole coupling torque of $\expval{\v L}$. Figure~\ref{fig:fig_5} shows that, without SOC, the relative directions of the dipole coupling torques of $\expval{\v L}$ and $\expval{\v S}$ are different. The spin magnetic torque contribution, $(-2\mu_B/\hbar)(\expval{\v S}\times \v B)_z = (-2\mu_B/\hbar)(\expval{S_x} B_y - B_x \expval{S_y})$ has the same direction as $d\expval{S_z}/dt$, since $B_y = 0$, and  $B_x < 0$ and $\expval{S_y} > 0$ throughout the motion (which can be seen in figure~\ref{fig:withoutSOC_B=500_a}). This causes the net direction of the spin magnetic torque contribution to be in the $-\u{z}$ direction. 
This difference is caused by the interaction of the electrons with the lattice, which prevents $\expval{\v L}$ from precessing as it would if the lattice were absent.

The third panel in figure~\ref{fig:fig_5c} shows the relative contributions of the torques shown in the previous two panels to the total electronic angular momentum, according to the terms in Eq.~\eqref{eq:J}. Since $\mean{\frac{d\expval{S_z}}{dt}}$ is much greater than any of the averaged torques related to the orbital angular momentum, the spin dominates the change in the total electronic angular momentum. As the large spin contribution to $\expval{J_z}$ is decoupled from the nuclei in the absence of SOC, the average torque experienced by the nuclei, $\mean{\Gamma}_{\textrm{int},z}$, is much smaller than $d\expval{J_z}/dt$.
At low $B$, the quasi-adiabatic limit is reached, and the torques reduce rapidly as the spin becomes unable to respond to the rate of rotation of the $B$ field (as was demonstrated in figures~2-4).

\subsubsection{\label{sec:low_field_with_SOC}With spin-orbit coupling}

The simulation results with SOC included are shown in figure~\ref{fig:fig_6}.
Figure~\ref{fig:fig_6a} shows the contributions to the torque that affect $\expval{L_z}$, along with the value of $\mean{d\expval{L_z}/dt}$ obtained by summing them.
For values of $B$ large enough to produce quasi-adiabatic results ($B \gtrapprox \SI{100}{T}$), the averaged dipole coupling torque and the averaged SOC torque are approximately independent of $B$. This is because the magnitude of $\expval{\v L}$, which is proportional to $B$ in the absence of SOC, is now determined primarily by the $\frac{\xi}{\hbar^2} \expval{\v{L} \cdot \v{S}}$ term in the Hamiltonian and no longer rises significantly as $B$ rises. As far as the orbital angular momentum is concerned, the mean spin $\expval{\v S}$, which is finite even when the applied magnetic field is zero, acts like a large magnetic field. The averaged interaction torque increases with increasing $B$, which leads to an increase in the overall orbital torque on the electrons, $\mean{\frac{d\expval{L_z}}{dt}}$.
For values of $B< 100$~T, the breakdown of quasi-adiabaticity is apparent, leading to a gradual increase in the magnitudes of the torques due to the SOC and orbital moment terms, and then a rapid decrease in all torques as $B$ becomes so small that the angular momenta are no longer able to follow as it changes direction. 

Several qualitative differences are apparent when comparing figure~\ref{fig:fig_6a} to its equivalent without SOC, figure~\ref{fig:fig_5a}. 
The SOC torque contribution, shown in red, is of course non-zero only when the SOC parameter $\xi$ is non-zero.
In addition, the orbital-magnetic torque term $(-\mu_B/\hbar)(\expval{\v L}\times \v B)_z$ is reversed in direction in figure~\ref{fig:fig_6a} in comparison to figure~\ref{fig:fig_5a}, and points in the same direction as the spin-magnetic torque term, $(-2\mu_B/\hbar)(\expval{\v S}\times \v B)_z$, when SOC is included. This can be understood as being due to SOC ensuring that $\expval{\v L}$ and $\expval{\v S}$ are more closely aligned, which changes the sign of $\expval{L_y}$ (which can be seen by comparing figures~\ref{fig:withoutSOC_B=500_a} and~\ref{fig:withSOC_B=500_1}).
In the presence of SOC, the torque on the nuclei due to the electrons, $-\mean{\Gamma}_{\textrm{int},z}$, is many times larger than in its absence. For example, at $B = \SI{100}{T}$, the difference is a factor of approximately 15.


In the presence of SOC, the magnitude of $\expval{\v L}$ is much larger due to its coupling to $\expval{\v S}$, which is large due to Stoner exchange. As a result, the terms in figure~\ref{fig:fig_6a} can be large for small $B$, which causes a large $-\Gamma_{\textrm{int},z}$ even for small $B$. This allows for a torque on the nuclei due to the electrons to be observable for experimentally realistic $B$ field strengths. Now, since $\expval{\v L}$ is locked to $\expval{\v S}$ by the SOC and has a non-zero value even when $B = 0$, both $\expval{\v L}$ and $\expval{\v S}$ behave paramagnetically.


Figure~\ref{fig:fig_6b} shows the averaged torque contributions in the spin equation of motion. With SOC included, the average magnitude of $\frac{d\expval{S_z}}{dt}$ is approximately the same as it is in the absence of SOC. The SOC term is non-zero and takes the same sign as the magnetic dipole contribution to the change in spin angular momentum.

Figure~\ref{fig:fig_6c} shows the averaged torques from Eq.~\eqref{eq:J}. 
Comparing figures~\ref{fig:fig_5c} and~\ref{fig:fig_6c}, we see that the averaged torque on the electrons due to the externally applied field, $(\mean{\expval{\v \mu}\cross\v B})_z$, is approximately the same with and without SOC. By contrast the torque on the electrons due to the nuclei, $-\mean\Gamma_{\textrm{int},z}$ is greatly enhanced in the presence of SOC, in particular at low magnetic field strengths. Since the interaction torque, $-\mean\Gamma_{\textrm{int},z}$, is the torque acting on the nuclei due to the spin-lattice interaction, it can be thought of as the torque that enacts the Einstein-de Haas effect. As a result, figures~\ref{fig:fig_5c} and~\ref{fig:fig_6c} show that the Einstein-de Haas effect would not be observed for low magnetic field strengths if SOC was not present, and that the spin-lattice torque is significant at low field strength when SOC is included.

Since $\mean{\frac{d \expval{J_z}}{dt}}$ is the sum of the averaged torque on the electrons due to the external magnetic field and due to the nuclei, it is also increased in the presence of SOC. This makes sense when comparing the magnitudes of $\expval{\v L}$ and $\expval{\v S}$ at $t=0~\textrm{a.u.}$ in figures~\ref{fig:withoutSOC_B=500_a} and~\ref{fig:withSOC_B=500_1}, in which we see $\expval{\v S}(t=0)$ is approximately the same in both cases, while $\expval{\v L}(t=0)$ is enhanced in the presence of SOC, implying that a greater net change in total electronic angular momentum is required to reverse the sign of $\expval{\v J}$ since the average torque over the duration of the simulation is given by $-2\expval{\v J}(t=0) / T_f$.
Figure~\ref{fig:fig_6c} shows that when the orbital and spin contributions to the torque are summed, it is clear that the torque on the \Fe{15} cluster, $\Gamma_{\textrm{int},z}$, is a fraction of the torque exerted on the system by the externally applied $B$ field.

\section{Conclusions}\label{sec:conclusions}


This work set out to investigate the quantum mechanical origins of the Einstein-de Haas effect in a \Fe{15} cluster from first principles, by means of simulation using a non-collinear TB model.
It was shown that in a slowly rotating $B$ field, orbital and spin angular momenta can reverse their orientations, leading to a measurable torque on the cluster.
Despite the computational challenge of reaching physically realistic timescales for the simulation, the qualitative features of the evolution can be extrapolated by use of the adiabatic theorem. However, the net transfer of angular momentum to the iron is a non-adiabatic process.
By analysing the trends of the torque contributions as $B$ becomes small, it has been verified that SOC greatly enhances the interaction torque on the Fe\textsubscript{15} cluster due to coupling to the Stoner exchange-induced spin moment. The enhancement due to SOC is especially pronounced for low magnetic field strengths. 
This work demonstrated a quantum mechanical model capable of simulating the Einstein-de Haas effect in a ferromagnetic cluster and revealed the physical mechanisms which drive the timescales causing this effect. 
\\\\
The tight-binding model employed in this work is not gauge invariant, thus a significant improvement would be to use London Orbitals~\cite{london} to remove any arbitrariness arising from the choice of gauge. 
Experimentally, it would be valuable to visualize the rotation of the spin of a single ferromagnetic domain in order to confirm or deny whether the spin passes through $\expval{\v S} = \v 0$, or whether it follows a path closer to a rotation through an arc of a circle.
One speculative application of this work is that the tight binding model may be used to derive a macroscopic description of spin-lattice coupling effects using the formalism of molecular dynamics, which would make the spin-lattice dynamics of much larger systems tractable and relevant to engineering and industrial applications.

\ack
This work was supported through a studentship in the Centre for Doctoral Training on Theory and Simulation of Materials at Imperial College London, funded by EPSRC grant EP/L015579/1. This work has been carried out within the framework of the EUROfusion Consortium, funded by the European Union via the Euratom Research and Training Programme (Grant Agreement No 101052200 — EUROfusion) and was partially supported by the Broader Approach Phase II agreement under the PA of IFERC2-T2PA02.
Views and opinions expressed are however those of the author(s) only and do not necessarily reflect those of the European Union or the European Commission.
Neither the European Union nor the European Commission can be held responsible for them. 
 We acknowledge support from the Thomas Young Centre under grant TYC-101 and the RCUK Energy Programme Grant No. EP/W006839/1.

\appendix
\section{The Adiabatic Theorem specialized to the Einstein-de Haas Effect}\label{appendix:adiabatic_theorem}

This appendix shows how the adiabatic theorem may be manipulated into a form which enables the timescales for transitions between diabatic and adiabatic behaviour to be calculated. A variety of proofs of the adiabatic theorem exist in the literature~\cite{Born&Fock,Messiah,Bransden}, here we follow the approach of Griffiths~\cite{Griffiths}.

Instantaneous eigenstates of the time-evolving Hamiltonian are defined by,
\begin{align}
	H(t, \v m_a(t)) \ket{\psi_n(t)} = E_n(t) \ket{\psi_n(t)} \ n = 1,2,\dots\label{eq:instantaneous_eigenstates_def}
\end{align}
These eigenstates are not solutions of the time-dependent Schr\"{o}dinger equation in general. 

For a solution of the Schr\"{o}dinger equation, $\ket{\Psi(t)}$, consider a wavefunction which begins its evolution at $t=0$ in an energy eigenstate,
\begin{align}
	\ket{\Psi(0)} = \ket{\psi_m(0)}.
\end{align}
Expanding the wavefunction as a linear superposition of instantaneous energy eigenstates gives,
\begin{align}
	\ket{\Psi(t)} = \sum_n c_n(t) \ket{\psi_n(t)}.
\end{align}
Substituting this into the Schr\"{o}dinger equation, and left-multiplying by $\bra{\psi_m(t)}$ yields
\begin{align}
	i\hbar\dot c_m &= \bigg(E_m(t) - i\hbar\braket{\psi_m|\dot \psi_m}\bigg) c_m - i\hbar \sum_{n\neq m} \braket{\psi_m|\dot \psi_n} c_n. \label{eq:adiabatic_theorem}
\end{align}

Taking the time derivative of the instantaneous eigenstates as defined in Eq.~\eqref{eq:instantaneous_eigenstates_def} informs us that for $m \neq n$
\begin{align}
	\braket{\psi_m | \dot \psi_n(t)} &= \frac{(\dot H)_{mn}}{E_n-E_m},\label{eq:17}
\end{align}
where $(\dot H)_{mn} = \braket{\psi_m|\dot H | \psi_n}$. Thus, Eq.~\eqref{eq:adiabatic_theorem} can be rewritten as
\begin{align}
	i\hbar\dot c_m &= \bigg(E_m(t) - i\hbar\braket{\psi_m|\dot \psi_m}\bigg) c_m - i\hbar \sum_{n\neq m}\frac{(\dot H)_{mn}}{E_n-E_m} c_n. \label{eq:adiabatic_theorem2}
\end{align}
For the system to remain in the ground state throughout the evolution, the coupling term must be small in order for the expansion coefficients of higher instantaneous energy eigenstates not to become significant. Thus the criterion for the quantum adiabatic approximation which must be satisfied for the system to remain adiabatic is:
\begin{align}
	\bigg\lvert \frac{(\dot H)_{mn}}{E_n-E_m} \bigg\rvert 
	\ll 1 \qquad \forall\ n\neq m.
\end{align}
The couplings can be grouped depending on the magnitude of the splitting $\lvert E_n - E_m \rvert$. The smallest differences in energy come from states which are split by $-\v\mu\cdot\v B$, which yields the condition
\begin{align}
	T_{f} > \frac{2\pi\hbar}{\Delta E_{-\v \mu \cdot \v B}},
\end{align}
where $T_{f}$ is the duration of the simulation. SOC causes larger energy level splittings, with an associated timescale given by
\begin{align}
	T_{f} > \frac{2\pi\hbar}{\Delta E_{SOC}}.
\end{align}
The largest energy splittings in the Hamiltonian are due to $H_0$, which lead to a separate timescale,
\begin{align}
	T_{f} > \frac{2\pi\hbar}{\Delta E_{H_0}}.
\end{align}

\nocite{*}

\bibliographystyle{iopart-num}
\bibliography{bibliography}

\end{document}